\documentclass[preprint,showpacs,preprintnumbers,amsmath,amssymb,superscriptaddress]{revtex4-1}
 \usepackage{graphicx}
 \usepackage{dcolumn}
 \usepackage{bm}
 \usepackage{ulem}
\usepackage{tensor}
\usepackage{xfrac}

\def\OOmega{{\bm \Omega}}
\def\r{{\bm r}}
\def\z{{\bm z}}

\def\b{{\bm b}}
\def\rrho{{\bm \rho}}
\def\l{{\bm l}}
\def\s{{\bm s}}
\def\q{{\bm q}}

\begin{document}

\title{Neutron Removal from the Deformed Halo \mbox{$^{31}$Ne} Nucleus}
\author{Juhee Hong}
\affiliation{\mbox{Rare Isotope Science Project,
Institute for Basic Science,} \\
Daejeon 305-811, Korea}
\author{C.A. Bertulani}
\affiliation{\mbox{Department of Physics, Texas A\&M University-Commerce},
\\
Commerce, TX 75429, USA
}
\author{A.T. Kruppa}
\affiliation{\mbox{Institute for Nuclear Research, Hungarian Academy of Sciences},
\\
Debrecen, PO Box 51, H-4001, Hungary
}

\date{\today}

\begin{abstract}
Experimental data on Coulomb breakup and neutron removal indicate that 
\mbox{$^{31}$Ne}  is one of the heaviest halo nuclei discovered so far.  
The possible ground state of \mbox{$^{31}$Ne} is either $3/2^-$ coming from 
p-wave halo or $1/2^+$ from s-wave halo. 
In this work, we  develop a treatable model to include deformed wave functions and a dynamical 
knockout formalism which includes  the dependence on the nuclear orientation
to study   the neutron removal from \mbox{$^{31}$Ne} projectiles at
energies around $E\approx 200$ MeV/nucleon. A detailed account of the effects of deformation
on cross sections and longitudinal momentum distributions is made.
Our numerical analysis indicates a preference for the \mbox{$^{31}$Ne}  
ground state with spin parity $3/2^-$.  
\end{abstract}

\maketitle

\section{Introduction}

The neutron-rich \mbox{$^{31}$Ne} nucleus is expected to be strongly deformed with the 
pf-intruder configuration near the island of inversion. Its halo structure  is relevant because it is
one of the heaviest halo nuclei discovered so far \cite{exp1}. The known data  seem to indicate a 
mixing of valence spherical orbitals due to deformation. 
The valence neutron of \mbox{$^{31}$Ne} is found to be in $2p_{3/2}$ 
or in $2s_{1/2}$, but not in 
$1f_{7/2}$ which is expected from the standard shell model 
\cite{ne9,ne10,ne16,ne17,ne18,ne22,ne19,ne20,ne21,Ne}. 
There are also uncertainties in the deformation and the neutron binding energy for two 
possible ground spin-parity states. 

In Ref. \cite{hamamoto}, the measured large Coulomb breakup cross section of 
\mbox{$^{31}$Ne} was interpreted simply  in terms of a p-wave neutron halo together with the deformed core. 
Hamamoto \cite{hamamoto} suggests that the ground state of \mbox{$^{31}$Ne} has either the spin-parity $1/2^+$ with a neutron separation energy $S_n>0.5$ MeV and 
quadrupole deformation $\beta_2\gtrsim0.59$ or the spin-parity
$3/2^-$ with $S_n<0.5$ MeV.  The $1/2^+$ assignment arises from the Nilsson level [200 1/2] while the $3/2^-$ assignment is due to either
(1) the [321 3/2] Nilsson level with $S_n<0.2$ MeV and $0.40\lesssim\beta_2\lesssim0.59$   
or (2) the [330 1/2] level with $0.2\, \mbox{MeV}<S_n<0.5$ MeV and $0.22\lesssim\beta_2\lesssim0.30$.    

In this work we consider one-neutron removal reactions as a probe of deformation and calculate cross sections and
longitudinal momentum distributions with respect to the incident beam direction. 
We use a modified version of the  Glauber model developed in Ref. \cite{bertsch}   
and extensively used in the literature \cite{tostevin_hansen,momdis,bertulani}.  
In this model, a few approximations are made. It is assumed that
the excitation energy of the relative motion between the core and the removed 
neutron is much smaller than the projectile energy and is neglected. 
This adiabatic condition is well satisfied for collisions on a light target with projectiles
at intermediate energies ($E_{\rm lab} \gtrsim 50$ MeV/nucleon). 
In the spectator-core approximation, the core can 
be at most elastically scattered by the target. 
With these assumptions, the shape of the momentum distribution of core can be used to determine the degree 
of deformation of projectiles, angular momentum content, and the binding energy of a valence nucleon. 
As an example, we will show that these distributions can be used to identify the spin-parity of the ground state of \mbox{$^{31}$Ne}. 

Theoretical studies of nucleon removal reactions from deformed projectiles have been 
reported previously \cite{zelevinsky,tostevin3,tostevin2,tostevin,Ne,singh} and
the longitudinal momentum distribution of  stripping reactions has been calculated
by using the Nilsson model.
Glauber-type deformed potential S-matrices have been used in 
Ref. \cite{zelevinsky}, and the core-target S-matrix has been calculated in 
the absorbing-disk approximation \cite{esbensen} in Ref. \cite{tostevin}. 
In this work, we calculate the momentum distribution with 
orientation-dependent S-matrices obtained by the nuclear ground state 
densities and the nucleon-nucleon cross section \cite{fnn,book,trr}.  Orientation dependence
is important because the valence nucleon knockout depends on the angle between the
intrinsic deformation axis and the beam axis. 
The deformed states used in our calculations are obtained by a solution of
coupled equations for a deformed Woods-Saxon potential. 
The approach is superior than the Nilsson model because the harmonic oscillator wave functions used 
in the model decay too fast at long distances and the asymptotic behavior of  the Nilsson states is not correct. 
This is of relevance for reactions induced by halo nuclei, such as $^{31}$Ne.

In Section \ref{deform}, we discuss nucleon removal reactions from deformed 
projectiles in an extension of the Glauber model of Ref. \cite{bertsch} to include deformation. 
We apply this formalism to study one-neutron removal reactions from deformed \mbox{$^{31}$Ne} in 
Section \ref{result}. 
The longitudinal momentum distributions and the total cross sections are then
considered as functions of quadrupole deformation and neutron binding energy. 
By including spherical calculations for neutron removal from the core to populate excited core states, the inclusive momentum 
distributions are compared with experimental data.  Finally, we summarize our results in Section \ref{summary}. 

\section{Nucleon Removal from Deformed Projectiles}
\label{deform}

We consider single-nucleon removal reactions from a two-body composite 
projectile consisting of a core and a valence nucleon.  In our model, the nucleon removal 
reactions have contributions from two processes: diffraction dissociation 
and stripping \cite{bertsch,HM85}. 
Diffraction dissociation is the elastic breakup process in which 
a valence nucleon is separated from the core whereas the target remains in its ground 
state. 
In the stripping or absorption process, the removed nucleon reacts with the 
target and the target is excited. 
In nucleon removal from halo nuclei, the momentum distribution becomes 
narrow due to the large spatial extension of their intrinsic wave functions. 
Thus, it is useful to interpret the distribution in terms of the momentum space 
wave function of the halo nuclei. 
Using the Glauber model \cite{bertsch,HM85},  the
nucleon removal under a spherical potential has been discussed by several  
authors, e.g., in Refs. \cite{tostevin_hansen,bertulani,momdis}. 
In our work, the momentum distribution and the cross sections of stripping 
and diffraction dissociation are calculated with an extended version of the numerical code 
MOMDIS \cite{momdis} to accommodate the changes described below. 
In the next subsections, we show how we include the projectile deformation in the nucleon removal 
cross sections in reactions with a  spherical target. 

\subsection{Deformed States}

To obtain the projectile deformed states, we use an updated version of the numerical code  
PSEUDO \cite{pseudo}. 
In the single-particle model for a deformed potential with axial symmetry, we have
\begin{equation}
V(\r,\hat{\OOmega})=\sum_{\lambda=0,2,4,\cdots}V_\lambda(r)Y_{\lambda 0}(\hat{\OOmega}) \, ,
\end{equation} 
and the ground state can be written as 
\begin{equation}
\Psi_{\omega}(\r,\hat{\OOmega})
=
\sum_{\alpha,m,s_z}\langle lm {1\over 2} s_z|j\omega\rangle
\frac{u_{\alpha\omega}(r)}{r}Y_{lm}(\hat{\OOmega}) 
\chi_{\frac{1}{2}s_z} \, ,
\end{equation}
where $\alpha=\{l,j\}$ and $\hat{\OOmega}$ defines the orientation of the 
symmetry axis relative to the laboratory system. 
The radial wave function is obtained by solving the following coupled system 
of ordinary differential equations:
\begin{equation}
-\frac{\hbar^2}{2\mu}\left[\frac{d^2}{dr^2}-\frac{l_\alpha(l_\alpha+1)}{r^2}\right]
u_{\alpha\omega}(r)
+
\sum_{\alpha',\lambda}\left[V_{\alpha\alpha'\lambda}(r)+
\frac{1}{r}V_{\alpha\alpha'\lambda}^{\rm sing}(r)\right]u_{\alpha'\omega}(r)
=Eu_{\alpha\omega}(r) \, ,
\end{equation}
where $V_{\alpha\alpha'\lambda}$ is the potential component corresponding to the coupling between $\alpha=\{l,j\}$ and $\alpha'=\{l',j'\}$ channels in the presence of deformation $\beta_{\lambda}$, and $V_{\alpha\alpha'\lambda}^{\rm sing}(r)/r$ corresponds to the singular part 
(such as spin-orbit interaction) of the potential which requires a special numerical treatment. In the present work, we consider quadrupole deformation ($\lambda=2$) only.  
 
The potential consists of the nuclear potential and the Coulomb potential. 
By expanding a deformed Woods-Saxon form factor with 
$R(\hat{\OOmega})=R_0[1+\sum_\lambda \beta_\lambda Y_{\lambda 0}(\hat{\OOmega})]$ 
($\beta_\lambda$ is the deformation parameter) 
and keeping only linear orders of deformation, 
the nuclear potential is given by 
\begin{equation}
V_N(\r,\hat{\OOmega})=-V_0f(r)-V_{SO}\left(\frac{\hbar}{m_\pi c}\right)^2\frac{1}{r}
\frac{df(r)}{dr} \, \l\cdot\s
+V_0R_0\frac{df}{dr}\sum_\lambda\beta_\lambda Y_{\lambda0}(\hat{\OOmega}) \, ,
\end{equation}
where $f(r)$ is a spherical Woods-Saxon form factor and 
$\hbar/(m_\pi c)=1.414$ fm is the pion Compton wavelength. 
The potential depths are adjusted to reproduce the ground state energy. 
The Coulomb potential is parameterized as
\begin{multline}
V_C(\r,\hat{\OOmega})=\frac{Z_1Z_2e^2}{r}\theta(r-R_C)+\frac{Z_1Z_2e^2}{2R_C}\left(3
-\frac{r^2}{R_C^2}\right)\theta(R_C-r) 
\\
+\sum_\lambda\frac{3Z_1Z_2e^2}{2\lambda+1}\left[
\frac{R_C^\lambda}{r^{\lambda+1}}\theta(r-R_c)
+\frac{r^\lambda}{R_C^{\lambda+1}}\theta(R_C-r)\right]\beta_\lambda 
Y_{\lambda 0}(\hat{\OOmega}) \, ,
\end{multline}
where $R_c$ is the Coulomb radius and $\theta(r)$ is the unit step function. 
For the case of $^{31}$Ne ($^{30}$Ne + n), the Coulomb potential has no influence in the calculations.
For more details, see Ref. \cite{pseudo}.

The basis functions are expressed in the projectile body-fixed frame with the 
$\hat{\z'}$-axis along the core symmetry axis. 
Thus, we need to project them (with $\omega'$) on the laboratory coordinate system with the 
$\hat{\z}$-axis along the beam direction 
\begin{equation}
\Psi_{\omega}(\r,\hat{\OOmega})
=
\sum_{\omega'}D_{\omega'\omega}^j(\hat{\OOmega}) \, \Psi_{\omega'}(\r) \, ,
\end{equation}
where $D_{\omega'\omega}^j(\hat{\OOmega})$ is the Wigner D-matrix with 
the Euler angles $\hat{\OOmega}=(\phi_o,\theta_o,0)$. 

\subsection{Reaction S-matrix}

In the eikonal approximation, the S-matrix is given by 
$S(\b)=\exp[i\chi(\b)]$ with 
\begin{equation}
\chi(\b)=-\frac{1}{\hbar v}\int dz \, V(\b+z\hat{\z}) \, , 
\end{equation}
where $v$ is the beam velocity along the $\hat{\z}$ axis, and 
$V(\b+z\hat{\z})$ is the optical potential for core-target or 
nucleon-target interaction. 
The eikonal phase is obtained from the nuclear ground state densities \cite{book,trr} as
\begin{equation}
\chi(\b,\hat{\OOmega})=\frac{1}{2\pi k_{NN}}\int d^3\r d^3\r' \, \rho_p(\r,\hat{\OOmega}) \rho_t(\r')
\int d^2\q \, f_{NN}(\q) \, e^{-i(\b-\rrho-\rrho')\cdot\q} \, ,
\end{equation}
where $k_{NN}$ is the nucleon-nucleon collision wave number, $\rho_p(\r)$ [$\rho_t(\r')$] is the nuclear density of the projectile 
[target], 
and $f_{NN}(\q)$ is the high energy nucleon-nucleon scattering 
amplitude at forward angles. 
Assuming a spherical projectile or orientation-independent S-matrix, 
the eikonal phase is 
\begin{equation}
\label{sphchi}
\chi_{\rm sph}(b)= {1\over k_{NN}}
\int dq \, q\rho_p(q)\rho_t(q)f_{NN}(q)J_0(qb) \, ,
\end{equation}
where we have taken the Fourier transform of the densities.  

For a deformed core, we expand the core density to linear orders of 
deformation \cite{defchi} 
\begin{equation}
\label{denexp}
\rho_c(\r,\hat{\OOmega})=\rho_c(r)+R_0\sum_{\lambda}\beta_\lambda Y_{\lambda 0}(\hat{\OOmega})
\frac{\partial\rho_c}{\partial r}\bigg\vert_{\beta_\lambda=0} \, .
\end{equation}
Here, the spherical harmonics need to be rotated into the laboratory frame, 
$Y_{\lambda 0}(\hat{\OOmega})=\sum_m D_{m0}^{\lambda}(\hat{\OOmega})
Y_{\lambda m}(\hat{\r})$. 
The eikonal phase of the core-target S-matrix is then given by 
\begin{multline}
\label{chidef}
\chi_{\rm def}(b,\hat{\OOmega})={1\over k_{NN}}
\int dq \, q\rho_c(q)\rho_t(q)f_{NN}(q)J_0(qb) 
\\ 
+\sum_{\lambda,m}R_0\beta_\lambda D_{m0}^\lambda(\hat{\OOmega}) \int d^3\r \, 
Y_{\lambda m}(\hat{\r}) 
\frac{\partial \rho_c}{\partial r}\bigg\vert_{\beta_\lambda=0}
\frac{1}{2\pi k_{NN}} \int d^2\q \, 
\rho_t(q)
f_{NN}(q)e^{-i(\b-\rrho)\cdot\q} \, . 
\end{multline}
For quadrupole deformed core, we obtain (see Appendix A) 
\begin{eqnarray}
\label{deformchi}
\chi_{\rm def}(b,\hat{\OOmega})&=& 
\frac{1}{k_{NN}}\int dq \, q\rho_c(q)\rho_t(q)f_{NN}(q)J_0(qb)
\nonumber\\
&&+\frac{\sqrt{5\pi}}{k_{NN}}R_0\beta_2 D_{00}^2(\hat{\OOmega})\int dr 
\frac{\partial\rho_c}{\partial r}\bigg\vert_{\beta_2=0} 
\nonumber\\
&&\qquad\times\int dq J_0(qb)\rho_t(q)f_{NN}(q)
\frac{1}{q^2r}\Big[(3-q^2r^2)\sin(qr)-3qr\cos(qr)\Big] 
\nonumber\\
&&+\sqrt{\frac{15\pi}{2}}\frac{1}{k_{NN}} R_0\beta_2 
[D^2_{20}(\hat{\OOmega})+D^2_{-20}(\hat{\OOmega})]
\int dr \,  \frac{\partial \rho_c}{ \partial r}\bigg\vert_{\beta_2=0}
\nonumber\\
&&\qquad\times
\int dq \, J_2(qb) \rho_t(q)f_{NN}(q) 
\frac{1}{q^2r}\Big[(3-q^2r^2)\sin(qr)-3qr\cos(qr)\Big] \, .
\end{eqnarray}

The core and target densities are obtained using  the liquid-drop model \cite{My70}. 
We also include the effect of the nucleon size. For a valence nucleon, we use a Gaussian form of density given by 
$\exp[-(r/0.7)^2]$. 
The nucleon-nucleon scattering amplitude, $f_{NN}(q)$, is parameterized as \cite{fnn}
\begin{equation}
f_{NN}(q)=\frac{k_{NN}\sigma_{NN}}{4\pi}(i+\alpha_{NN})e^{-\beta_{NN}q^2} \, , 
\end{equation}
where $\sigma_{NN}$, $\alpha_{NN}$, and $\beta_{NN}$ are obtained from fitting the 
nucleon-nucleon scattering data. 

\subsection{Cross Sections}

The basic assumption in the application of the Glauber theory to stripping reactions is that
one can write the cross sections as integrals over the transverse coordinates, 
and the impact parameter dependent S-matrix can be interpreted as a 
survival probability \cite{HM85}.  In nucleon removal reactions, the core reaches a detector intact 
(with the probability $|S_c|^2$) and the valence nucleon is absorbed by the target (with $1-|S_n|^2$, where $S_c$ and $S_n$ are the core-target and nucleon-target 
S-matrices, respectively) \cite{HM85,bertsch}. 

For unpolarized projectile beams, we need to average over all orientations. 
Thus, the longitudinal momentum distribution of the stripping cross section is given by \cite{bertsch}
\begin{equation}
\label{longdef}
\frac{d\sigma_{\rm str}}{dk_z}
=\frac{1}{2\pi}\frac{1}{4\pi}\int d\hat{\OOmega}\int d^2\b_n \,
[1-|S_n(\b_n)|^2]\int d^2\rrho \, 
\bigg\vert \int dz \, e^{-ik_zz} S_c(\b_c,\hat{\OOmega})
\Psi_{\omega}(\r,\hat{\OOmega}) \bigg\vert^2 \, . 
\end{equation}
The total cross section of stripping is calculated by integrating over $k_z$, yielding 
\begin{equation}
\sigma_{\rm str}=\frac{1}{4\pi}\int d\hat{\OOmega} 
\int d^2\b_n \, [1-|S_n(\b_n)|^2]
\int d^3\r \, \Psi_{\omega}^*(\r,\hat{\OOmega})| S_c(\b_c,\hat{\OOmega})|^2
\Psi_{\omega}(\r,\hat{\OOmega}) \, .\label{s2}
\end{equation}
Another process, elastic breakup or diffraction dissociation, can also be interpreted in terms
of survival amplitudes with help of the eikonal S-matrices \cite{bertsch}. Including the effects of deformation,
the total cross section for diffraction dissociation is 
\begin{multline}
\sigma_{\rm diff}=\frac{1}{4\pi}\int d\hat{\OOmega} 
\int d^2\b_c
\bigg[
\int d^3\r \, \Psi_{\omega}^*(\r,\hat{\OOmega})|S_n(\b_n) 
S_c(\b_c,\hat{\OOmega})|^2
\Psi_{\omega}(\r,\hat{\OOmega})
\bigg.
\\
\bigg. 
-\sum_{\omega'} \bigg\vert \int d^3\r \, \Psi_{\omega'}^*(\r,\hat{\OOmega})
S_n(\b_n)S_c(\b_c,\hat{\OOmega})\Psi_{\omega}(\r,\hat{\OOmega})\bigg\vert^2
\bigg] \, .\label{s3}
\end{multline}

One may first assume that the S-matrices do not depend on orientation and that deformation 
effects are solely due to the extended (halo) deformed single-particle wave functions. 
In this work, we also ignore interference contributions so that the coupled system 
is diagonal in the $\{l,j\}$ basis. 
Then the average over all orientations in the cross sections can be simplified by 
using the orthogonality of the Wigner D-matrices
\begin{equation}
\int d\phi \int d\theta \, \sin\theta \,  D_{kn}^{j*}(\phi,\theta,0) 
D_{k'n'}^{j'}(\phi,\theta,0)
=\frac{4\pi}{2j+1} \, \delta_{jj'} \, \delta_{kk'} \, \delta_{nn'} \, .
\end{equation}
This method allows us a straightforward use of the longitudinal momentum distribution and the cross 
sections of stripping and diffraction dissociation as calculated by the 
MOMDIS code for each $\{l,j\}$ component of the radial wave 
function and to sum the contributions separately, with the average over orientations
for the halo wave function accounted for properly. 

In general, the core-target S-matrix, $S_c(\b_c,\hat{\OOmega})$, depends on orientation. 
Therefore, we have modified MOMDIS to include deformation effects in the S-matrix or in the
eikonal phase as in Eq. (\ref{deformchi}). We  have then proceeded to calculate momentum distributions and cross sections for deformed projectiles following equations
(\ref{longdef})--(\ref{s3}). Our results are presented in the next section.

\section{Results and Discussions}
\label{result}

In this work we are particularly interested in neutron removal from $^{31}$Ne projectiles. We consider two possible ground spin-parity states, $3/2^-$ and $1/2^+$, for 
\mbox{$^{31}$Ne}. 
Following Ref. \cite{hamamoto,BohrMottelson}, we use the parameters of 
a deformed Woods-Saxon potential given by 
$R_0=R_{SO}=R_C=3.946$ fm, $a_0=a_{SO}=0.67$ fm, and $V_{SO}=-17.33$ MeV. 

\subsection{Deformed States in $^{31}$Ne}

\begin{figure}
\includegraphics[width=0.45\textwidth]{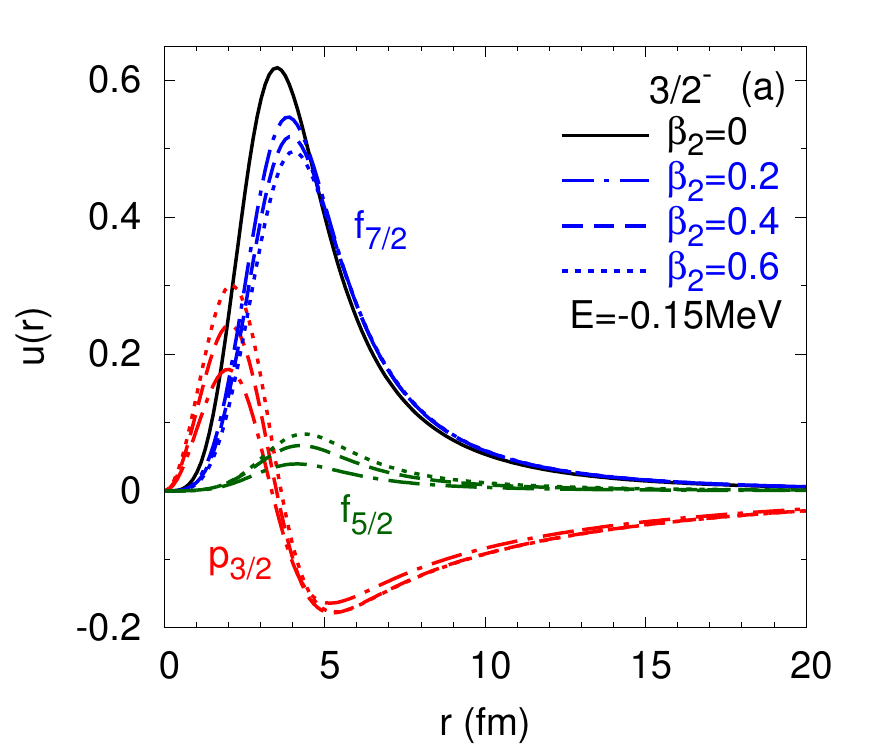}
\includegraphics[width=0.45\textwidth]{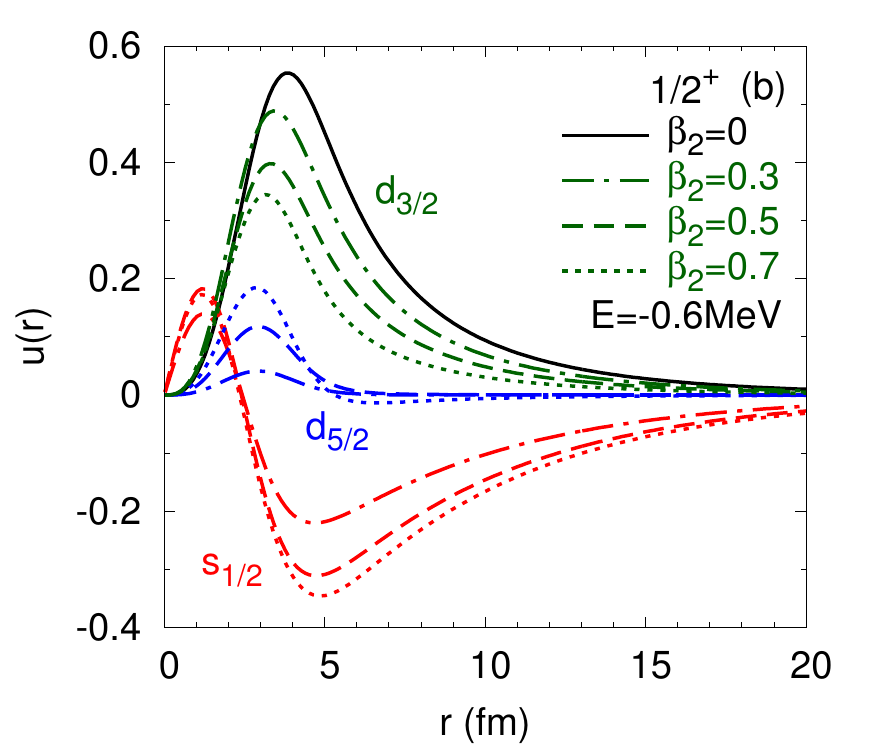}
\caption{ (Color online) 
(a) The channels $p_{3/2}$, $f_{5/2}$, and $f_{7/2}$ in $\Psi_{3/2^-}$. 
They are compared with the spherical single-particle radial wave 
function $1f_{7/2}$. 
(b) The channels $s_{1/2}$, $d_{3/2}$, and $d_{5/2}$ in $\Psi_{1/2^+}$. 
They are compared with the spherical single-particle radial wave 
function $1d_{3/2}$.  
} 
\label{nesol}
\end{figure}

For the quadrupole deformation $\beta_2=0.4$ and effective 
binding energy $E=-0.15$ MeV, we obtain the following normalized 
deformed state with spin-parity $3/2^-$: 
\begin{equation}
\label{nestate1}
\sum_\alpha u_{\alpha3/2^-}(r)=
\sqrt{0.24} \, p_{3/2}(r) 
+\sqrt{0.01} \, f_{5/2}(r) 
+\sqrt{0.75} \, f_{7/2}(r) 
\, ,
\end{equation}
with $V_0=-40.0$ MeV. 
In Fig. \ref{nesol} (a), the dashed lines represent the radial functions of 
the deformed state, 
and the solid line is the spherical single-particle radial wave function 
$1f_{7/2}$ obtained with $\beta_2=0$ and an appropriate potential depth $V_0$ to reproduce the same binding. We see that
the deformation shifts substantial contributions from the $1f_{7/2}$  to the $2p_{3/2}$ state, at the level of 24\%.
Since there are uncertainties in the quadrupole deformation parameter and the 
effective binding energy, we consider the deformed state for several values of 
$\beta_2$ and $E$. 
As the deformation decreases, the state $\Psi_{3/2^-}$ approaches the spherical 
wave function $1f_{7/2}$. It is worth noticing that it is mostly the tail of the wave function ($r\gtrsim 4$ fm) that contributes to
nucleon knockout reactions \cite{tostevin_hansen}. However, studies with tightly-bound nucleons have shown that this is not  always true
\cite{Gade2008}.

For $\beta_2=0.6$ and $E=-0.6$ MeV, another deformed state with the 
spin-parity $1/2^+$ can be obtained, 
\begin{equation}
\label{nestate2}
\sum_\alpha u_{\alpha1/2^+}(r)=
\sqrt{0.59} \, s_{1/2}(r) 
+\sqrt{0.37} \, d_{3/2}(r) 
+\sqrt{0.04} \, d_{5/2}(r) 
\, ,
\end{equation}
with the potential depth $V_0=-38.1$ MeV. 
In Fig. \ref{nesol} (b) we show that the state $\Psi_{1/2^+}$ approaches the spherical $1d_{3/2}$ state as 
$\beta_2\rightarrow0$. Deformation drains the contribution of the $1d_{3/2}$ state to $2s_{1/2}$ state and makes
their amplitudes nearly similar in strength.   

\subsection{Nucleon Knockout from $^{31}$Ne}

We now consider the single-neutron removal reaction 
\mbox{$^{12}$C}(\mbox{$^{31}$Ne},\mbox{$^{30}$Ne})X at the laboratory energy 
230 MeV per nucleon. 
The nucleon-nucleon scattering parameters $\alpha_{NN}=0.73$, $\beta_{NN}=0.58$, and 
$\sigma_{NN}=3.02 \, {\rm fm}^2$ are used \cite{trr,fnn}.  
The intrinsic matter density of the neutron (or proton), $\rho(r)$, is taken as
 a Gaussian function, corresponding to a form factor $\rho(q) = C \exp(-a^2q^2/4)$. 
 We use $a = 0.7$ fm for a nucleon density rms radius of 0.86 fm. 
The density rms radii of the core and target are $3.69$ fm and $2.90$ fm, respectively. 
For the core we have used a liquid-drop model density \cite{My70}, and for the carbon 
target we have used the density parametrization taken from Ref. \cite{DeVries}. We have verified that using a core density based on a Hartree-Fock-Bogoliubov 
calculation with the SLy5 Skyrme interaction does not change our results in a noticeable way. On the other hand, Ref. \cite{tostevin} (see their Fig. 5)  has 
shown  that cross sections have some sensitivity to the  relative sizes of the core and neutron wave function.

\begin{figure}
\includegraphics[width=0.45\textwidth]{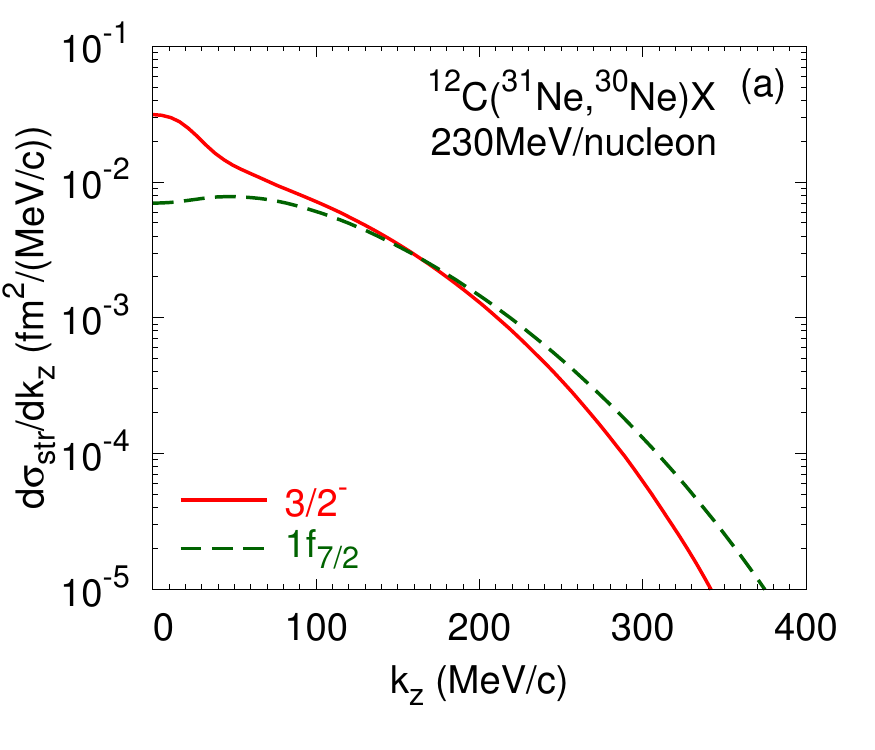}
\includegraphics[width=0.45\textwidth]{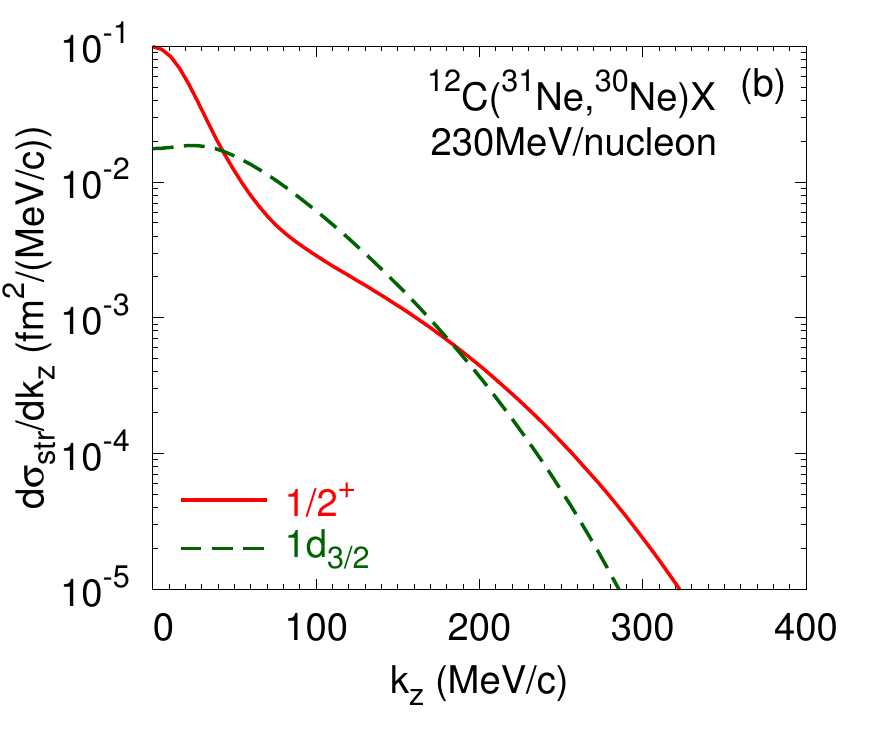}
\caption{ (Color online) 
The longitudinal momentum distribution for 
\mbox{$^{12}$C}(\mbox{$^{31}$Ne},\mbox{$^{30}$Ne})X at 230MeV/nucleon. 
(a) The solid line is for the deformed state $\Psi_{3/2^-}$ with $\beta_2=0.4$ and 
$E=-0.15$ MeV, and the dashed line is for the spherical single-particle wave 
function $1f_{7/2}$ with the same binding energy.  
(b) The solid line is for the deformed state $\Psi_{1/2^+}$ with $\beta_2=0.6$ and 
$E=-0.6$ MeV, and the dashed line is for the spherical single-particle wave 
function $1d_{3/2}$ with the same binding energy.  
}
\label{nelong} 
\end{figure}

In Fig. \ref{nelong}, we plot the calculated longitudinal momentum distributions of 
the two deformed states (solid lines) and  compare with the distributions obtained 
using the spherical single-particle wave functions (dashed lines). 
Near $k_z=0$, the cross sections obtained with the deformed states are larger 
than those obtained with the spherical wave functions. 
We note that the width of the momentum distribution changes with
 projectile deformation.  This is expected because the $p_{3/2}$ state and the 
$s_{1/2}$ state  
[in Eqs. \eqref{nestate1} and \eqref{nestate2}]
 have different, less space confining, centrifugal barriers than the corresponding $f_{7/2}$ and 
 $d_{3/2}$ states, respectively. Therefore, admixture with
 the $p_{3/2}$ state and the $s_{1/2}$ state will induce narrower momentum distributions due to a larger spatial extension
 of the wave functions.  In fact, the spatial extension (the rms radius) of the deformed states is larger than that of 
the spherical waves.  In summary,  we expect the deformed states to produce larger cross sections at low 
momentum and narrower momentum distributions in comparison with the spherical waves. 

\begin{figure}
\includegraphics[width=0.45\textwidth]{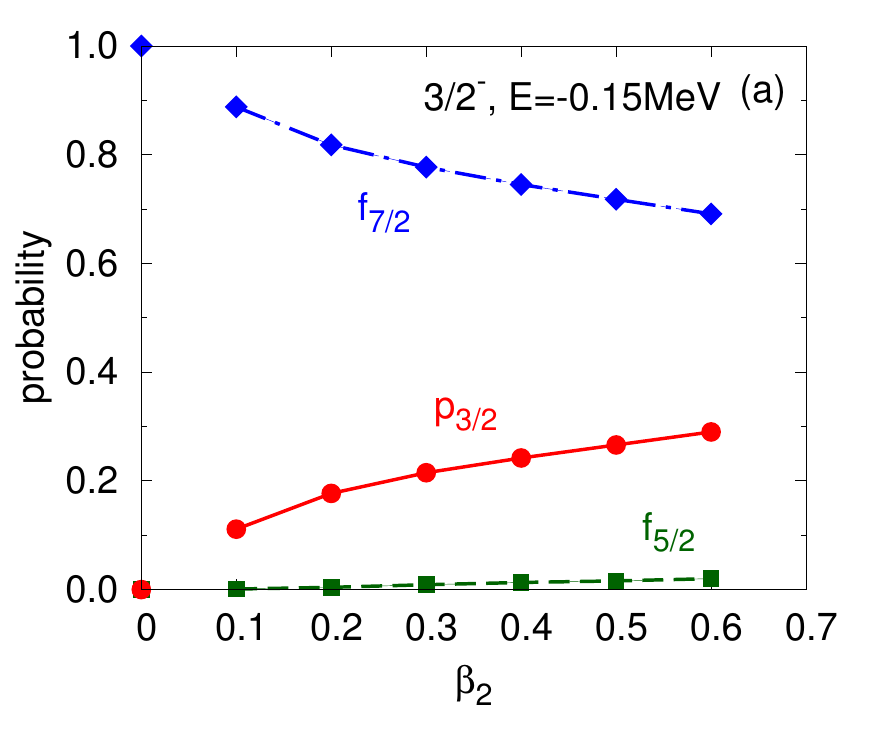}
\includegraphics[width=0.45\textwidth]{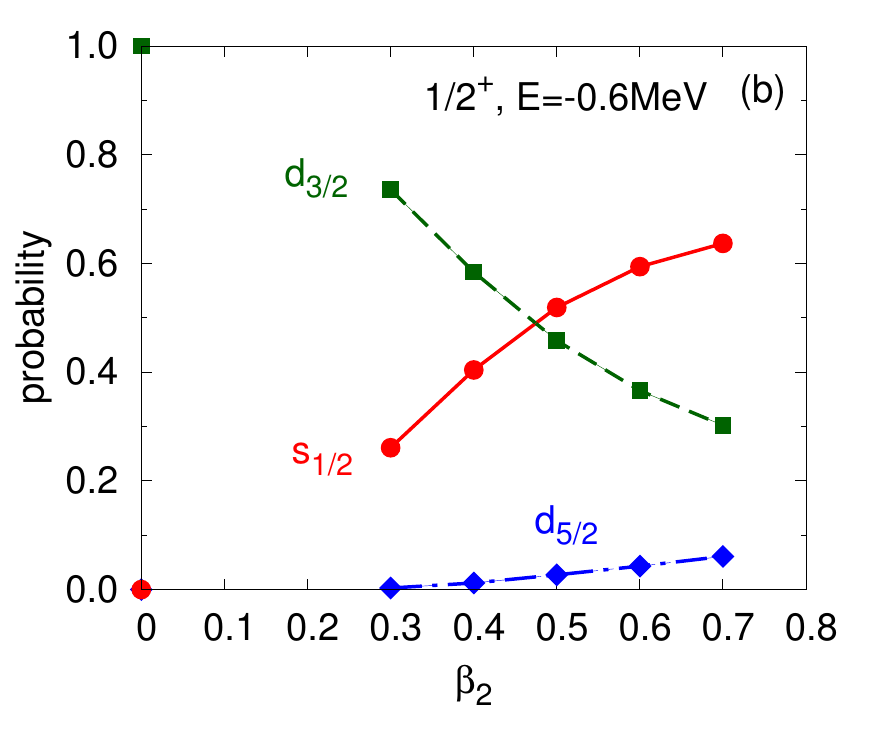}
\caption{
Squared amplitudes, $\int dr [u_{\alpha\omega}(r)]^2$, for the expansion of the 
states (a) $\Psi_{3/2^-}$ and (b) $\Psi_{1/2^+}$ in the basis $\alpha=\{l,j\}$.
}
\label{fgamp} 
\end{figure}

\begin{figure}
\includegraphics[width=0.45\textwidth]{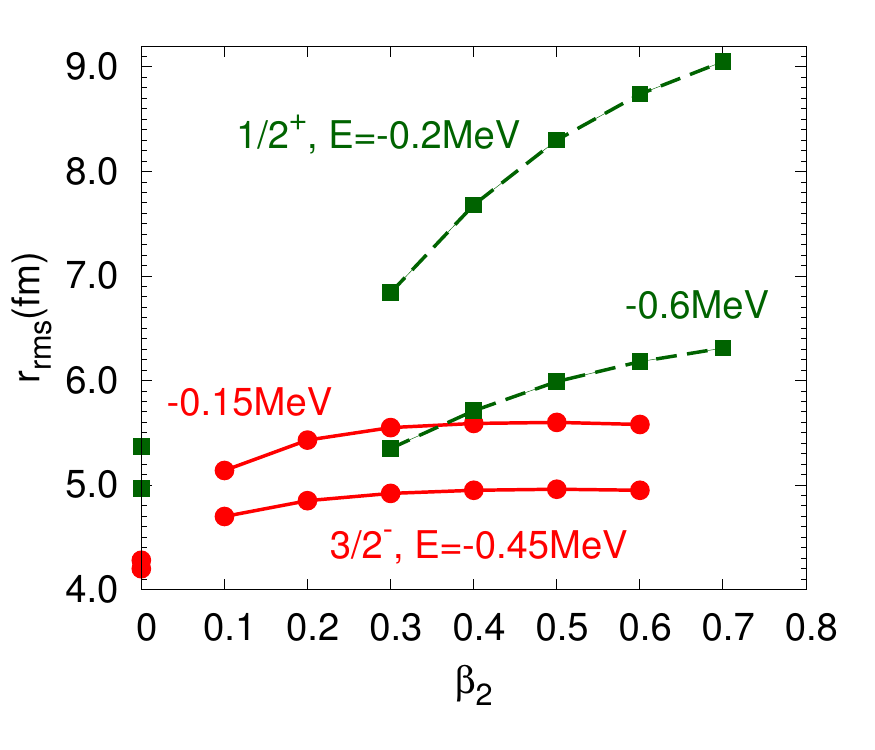}
\caption{ (Color online) 
The wave functions rms radii.}
\label{rms} 
\end{figure}

To investigate the core deformation effects, we have obtained the solutions 
with different  values of the deformation parameter $\beta_2$ for fixed energy $E$. 
All the parameters of the Woods-Saxon potential are fixed while the central potential 
depth is adjusted so that the energy $E$ of the state with $\beta_2$ is 
reproduced.  Depending on $\beta_2$, each shell occupation amplitude in Eqs. (\ref{nestate1}) and 
(\ref{nestate2}) changes and so do their wave functions,  as displayed in Fig. \ref{nesol} (see also Fig. \ref{fgamp}). 
The rms radius of the deformed states, 
$r_{\rm rms}=\left[\sum_{\alpha} \int dr \, [u_{\alpha\omega}(r)]^2r^2\right]^{1/2}$, 
is shown in Fig. \ref{rms}.  We note that $r_{\rm rms}$ increases with deformation for fixed $E$ for reasons explained above. 

\begin{figure}
\includegraphics[width=0.45\textwidth]{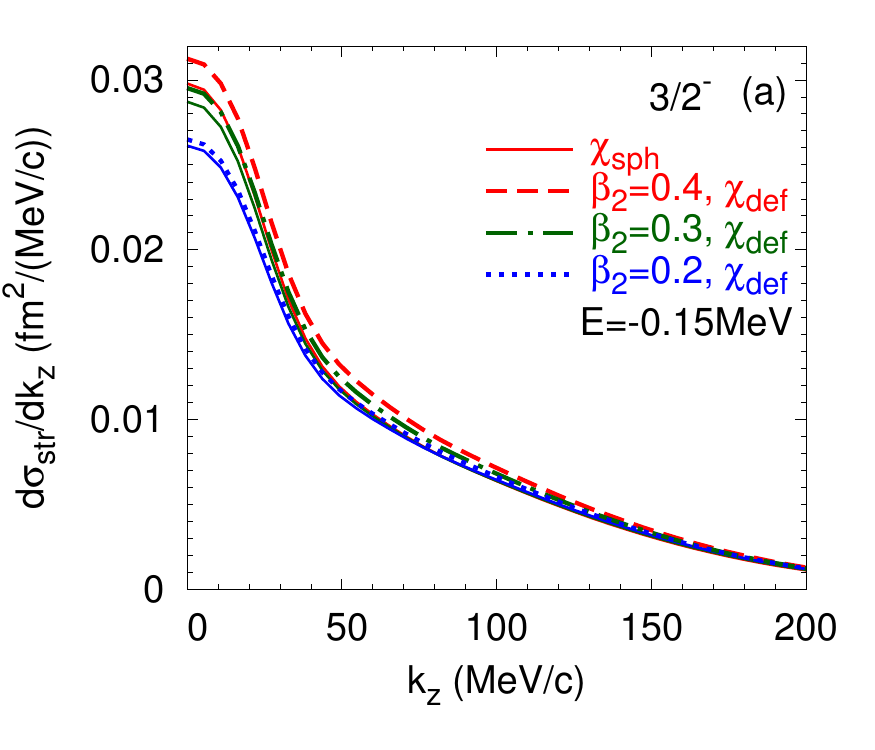}
\includegraphics[width=0.45\textwidth]{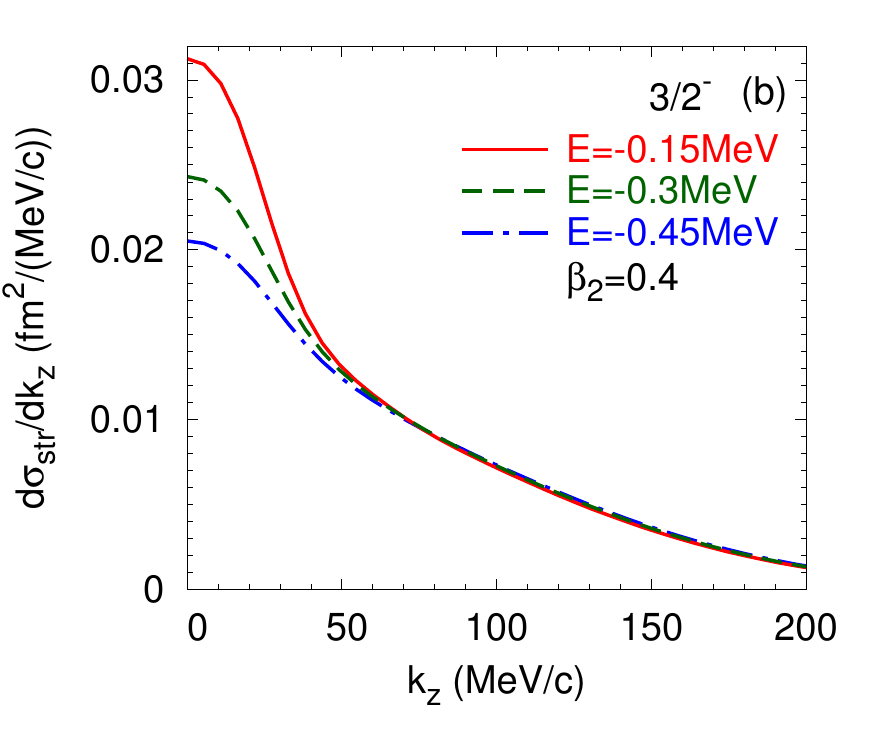}
\caption{ (Color online) 
Longitudinal momentum distributions of core fragments from neutron knockout from  the deformed state $\Psi_{3/2^-}$. 
(a) The dependence on the quadrupole deformation is displayed. 
The dashed, dash-dotted, and dotted lines represent the calculated 
results using Eq. (\ref{deformchi}). 
The solid lines are for the results obtained with Eq. (\ref{sphchi}). 
(b) The dependence on the effective binding energy of the valence neutron is shown.  
}
\label{ne1ds} 
\end{figure}

\begin{figure}
\includegraphics[width=0.45\textwidth]{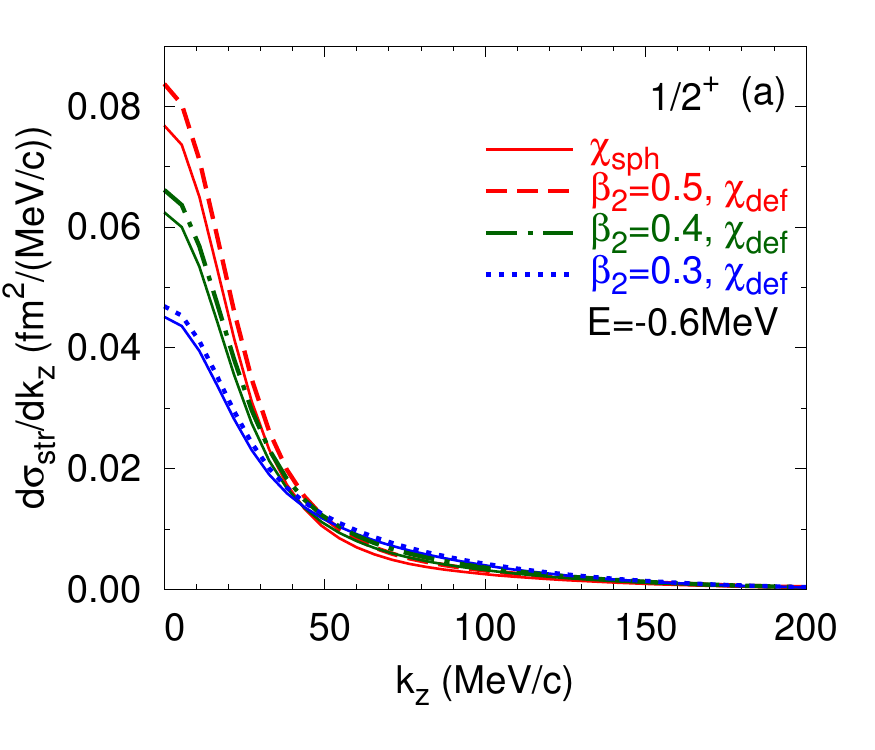}
\includegraphics[width=0.45\textwidth]{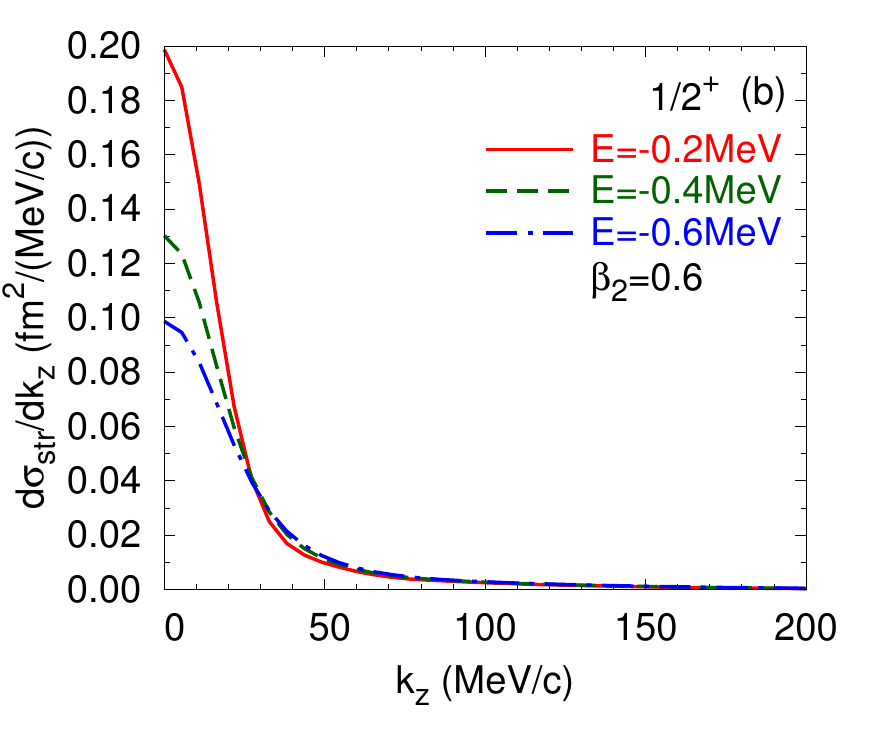}
\caption{ (Color online) 
Same as Fig. \ref{ne1ds} but for the deformed state $\Psi_{1/2^+}$. 
}
\label{ne2ds} 
\end{figure}

In Figs. \ref{ne1ds} (a) and \ref{ne2ds} (a), we present the dependence of the calculated 
longitudinal momentum distributions on projectile deformation. 
The calculated results using the orientation dependent core-target S-matrix, Eq. (\ref{deformchi}), 
are compared with the results using the spherical 
S-matrix, Eq. (\ref{sphchi}). 
With the deformed phase, the total cross section increases by $2-11\%$ compared with the spherical phase. 
The effect of the deformed phase is pronounced near $k_z=0$ for strong 
deformation.  
As expected from the relation between $r_{\rm rms}$ and deformation, 
Fig. \ref{rms}, the stronger the quadrupole deformation, the larger 
cross section we obtain. 

Similarly,  in Figs. \ref{ne1ds} (b) and \ref{ne2ds} (b), we have calculated the momentum distributions depending on $E$ for 
fixed $\beta_2$.  The
$r_{\rm rms}$ decreases as $|E|$ increases, which is reflected in the cross 
sections.  We observe that the widths of momentum distributions are sensitive to the 
effective binding energy of the valence neutron, as in the spherical case. Smaller widths are associated
with smaller binding due to the larger extension of the wave function. They are also influenced by the angular momentum $l$ content of the deformed state. 

\begin{figure}
\includegraphics[width=0.45\textwidth]{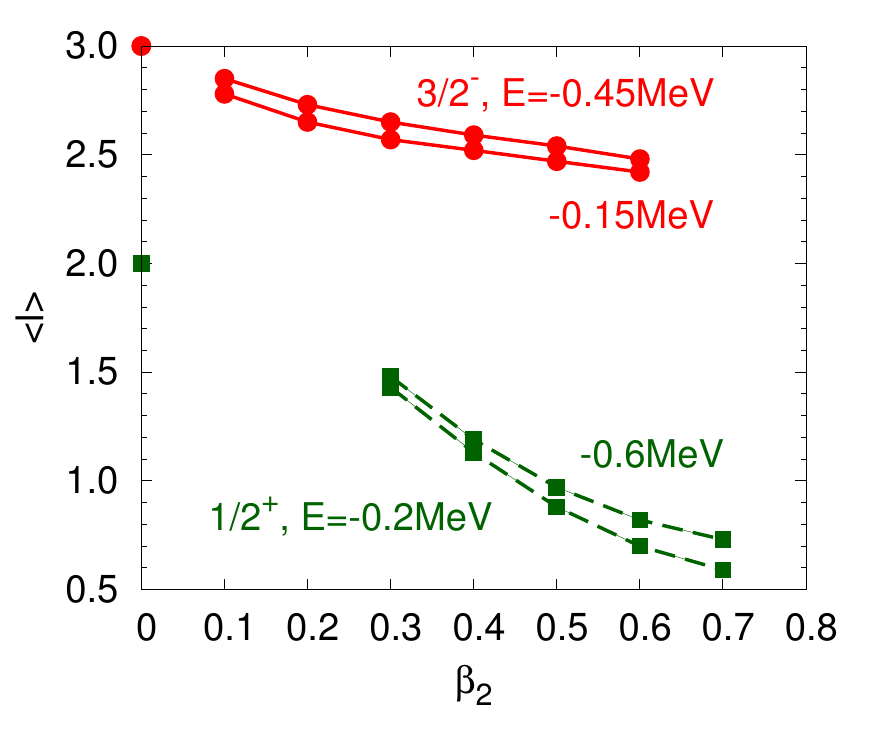}
\caption{ (Color online) 
The average $l$ values of the states $\Psi_{3/2^-}$ and $\Psi_{1/2^+}$. 
}
\label{ll} 
\end{figure}

Evidently, the cross sections increase with the rms radii of deformed states. 
This result contrasts with those reported in Ref. \cite{tostevin} where 
no correlation between the cross sections and the rms radii of the 
Nilsson states has been found.  
In Fig. \ref{ll}, we present the average $l$ value, 
$\langle l \rangle = \sum_{\alpha}l_\alpha \int dr [u_{\alpha\omega}(r)]^2 $, as a function of $\beta_2$ 
and $E$. 
As the quadrupole deformation grows, $\langle l \rangle$ decreases. 
This is because the probability of the $f_{7/2}$ [$d_{3/2}$] component in the 
state $\Psi_{3/2^-}$ [$\Psi_{1/2}^+$] 
decreases while that of $p_{3/2}$ [$s_{1/2}$] increases (see Fig. \ref{fgamp}). 
As the core mean field deformation changes, the occupation probabilities of spherical 
orbitals redistribute. 
Therefore, the cross sections and the widths of the corresponding momentum distributions 
change appreciably with deformation. 
When the binding energy grows, the probability of each channel changes in 
the opposite way.  
If the opposite behavior would  increase  $\langle l \rangle$ with $|E|$, then
the cross section would display an inverse trend with the average $l$ value. 
We note that the widths of longitudinal momentum distributions 
increase with the average $l$ value.  
Although our deformed states show different behaviors with the rms radii 
from the Nilsson states, the dependence of the cross sections and momentum 
distributions on $\langle l\rangle$ obtained with our method is similar 
to the results reported in Ref. \cite{tostevin}. 

\begin{figure}
\includegraphics[width=0.45\textwidth]{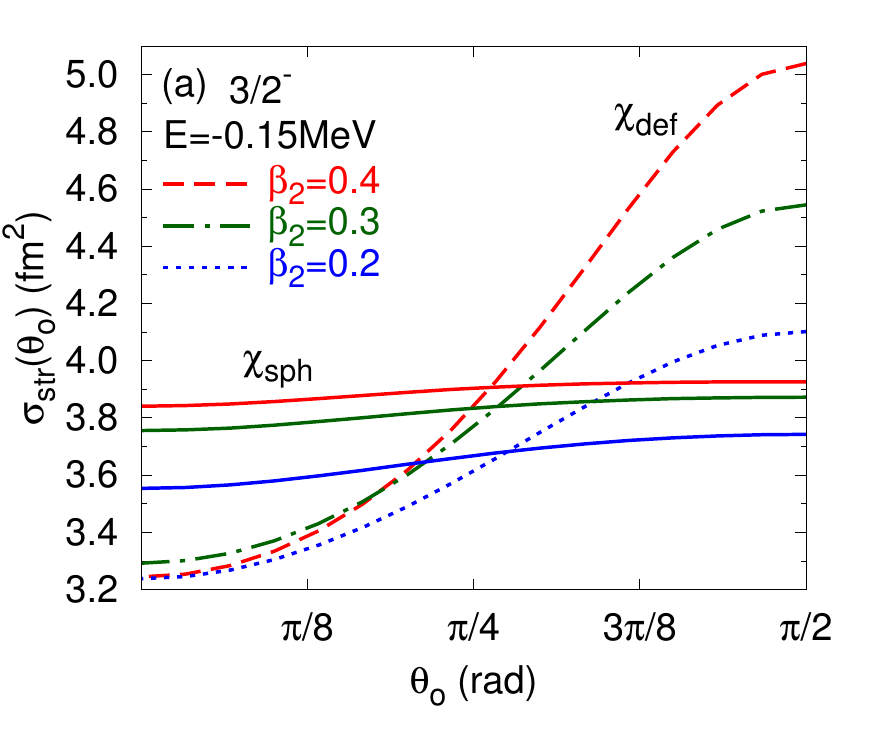}
\includegraphics[width=0.45\textwidth]{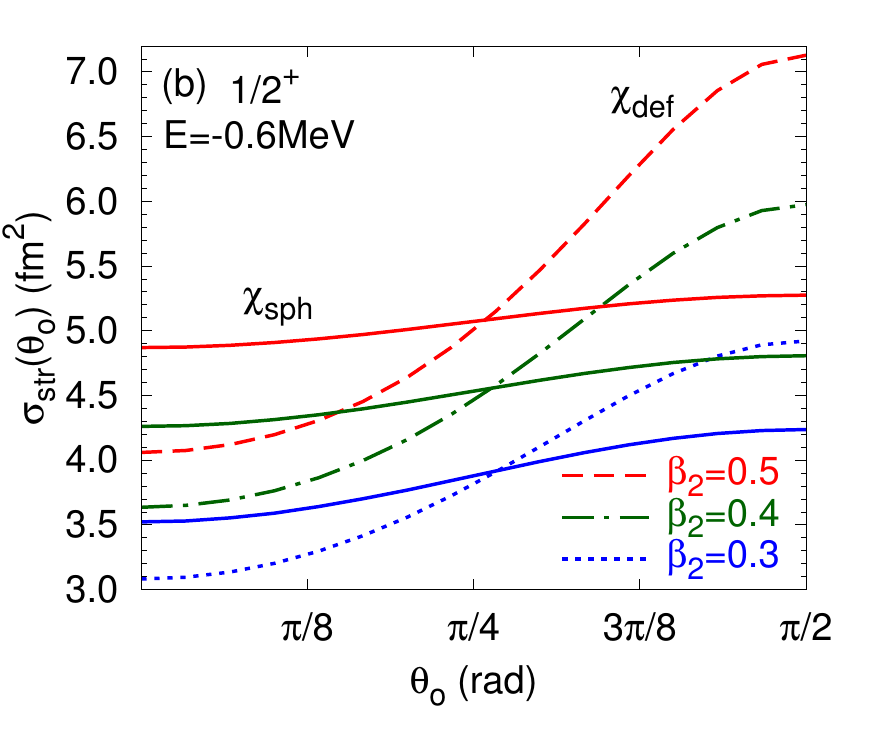}
\caption{ (Color online) 
Neutron knockout cross sections as a function of the Euler angle $\theta_o$ of the core symmetry 
axis for the states (a) $\Psi_{3/2^-}$ and (b) $\Psi_{1/2^+}$. 
The solid lines are calculated with Eq. (\ref{sphchi}), and the 
dashed, dash-dotted, and dotted lines are calculated with Eq. (\ref{deformchi}). 
}
\label{neang} 
\end{figure}

For prolate projectile deformation, the cross section is expected to be 
largest when the symmetry axis is perpendicular to the beam axis and 
smallest when parallel. 
This behavior is shown in Fig. \ref{neang} where we present our calculations for neutron removal  cross sections as a function of the 
Euler angle $\theta_o$ of the core symmetry axis. 
The distributions calculated with the deformed eikonal phase are 
compared with those with the spherical phase. 
In contrast to momentum distributions, we find that orientation distributions are sensitive to 
the orientation-dependence of the core-target S-matrix. 
Depending on deformation, the core density in Eq. (\ref{denexp}) changes with 
the angle $\theta_o$. 
The density has the maximum value at $\theta_o=\pi/2$ and minimum at $\theta_o=0$. 
Thus, the cross sections calculated with the deformed phase 
are larger [smaller] than the cross sections with the spherical 
phase near $\theta_o=\pi/2$ [$\theta_o=0$]. 
The deformation effects grow with $\beta_2$. 

\begin{figure}
\includegraphics[width=0.45\textwidth]{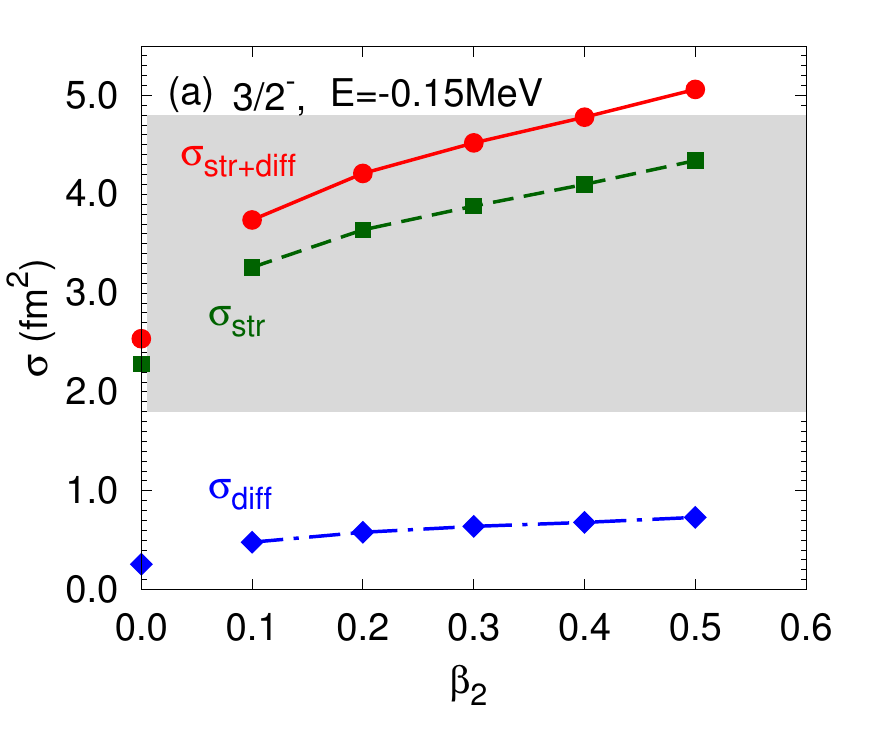}
\includegraphics[width=0.45\textwidth]{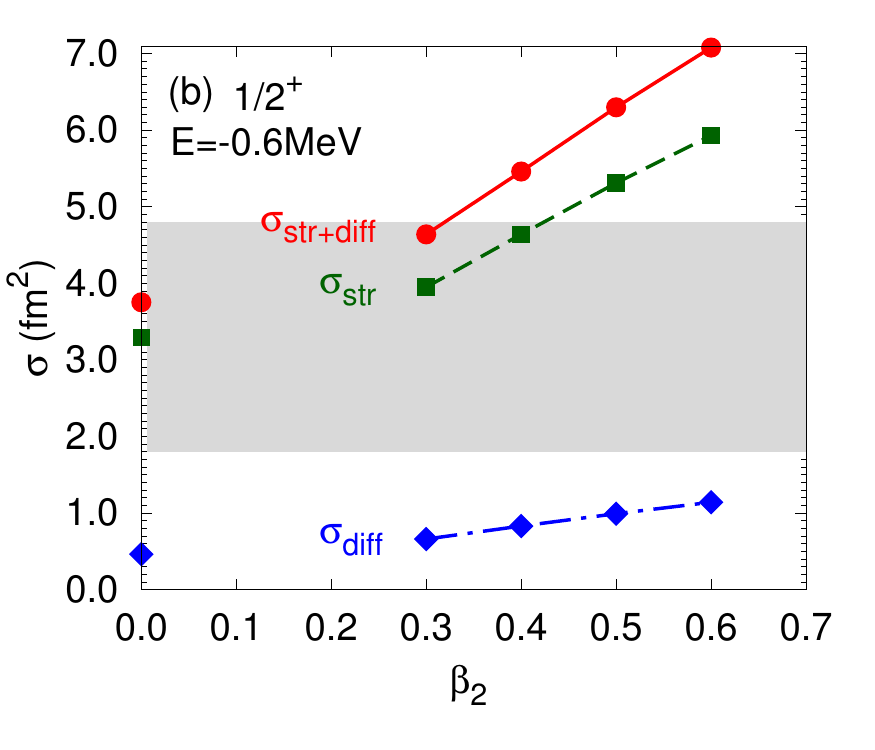}
\caption{ (Color online) 
Stripping, diffraction dissociation, and the sum of the two cross 
sections for the deformed states (a) $\Psi_{3/2^-}$ and (b) $\Psi_{1/2^+}$. 
The shaded region represents the measured cross section feeding the ground 
state of the residual core, 
$\sigma_{\rm exp}^{\rm gs}=3.3(1.5)$ ${\rm fm}^2$ 
\cite{exp2}. 
The experimental uncertainty is represented by the width of the shaded region. 
}
\label{netotd} 
\end{figure}

According to our calculations, Fig. \ref{netotd} shows the total cross sections of stripping and diffraction 
dissociation.  Both stripping and diffraction cross sections increase with 
quadrupole deformation.  The total cross section of diffraction dissociation amounts to $15-19\%$ of 
that of stripping. 

\begin{figure}
\includegraphics[width=0.45\textwidth]{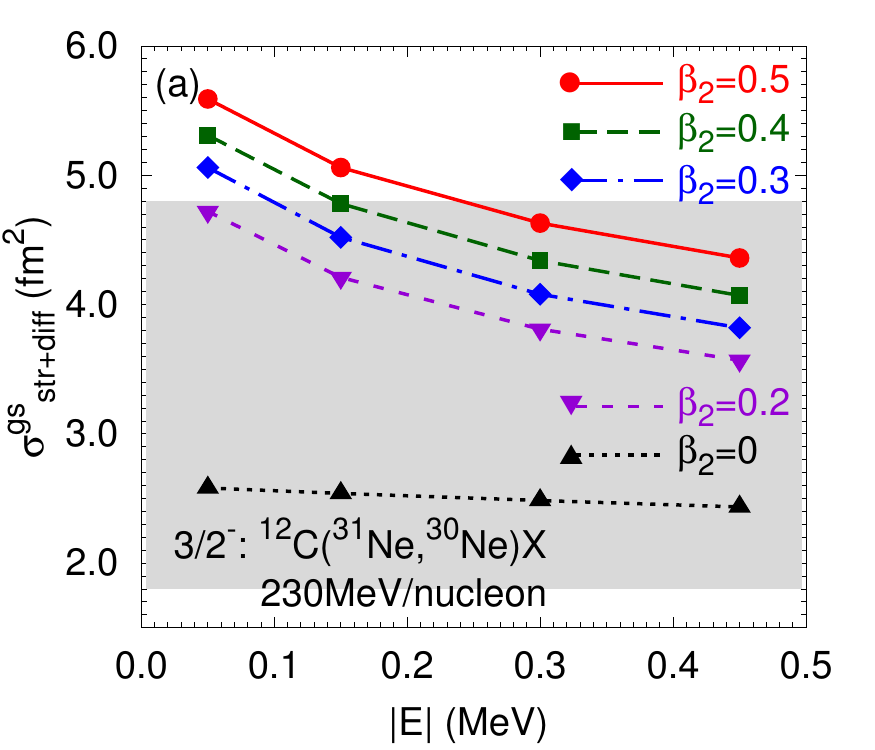}
\includegraphics[width=0.45\textwidth]{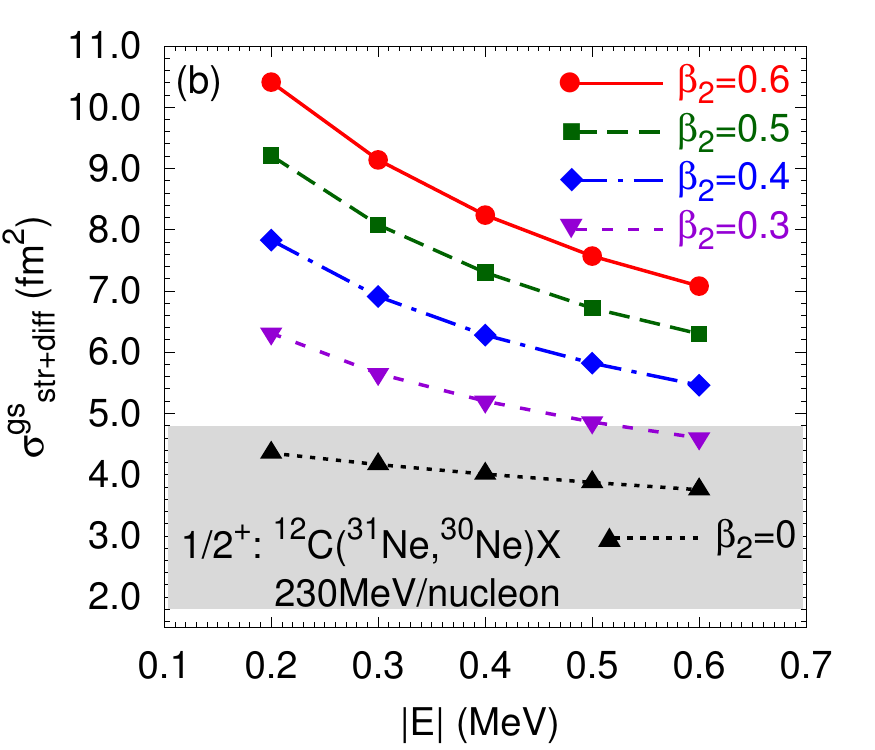}
\caption{ (Color online) 
The sum of stripping and diffraction cross sections for the deformed 
states (a) $\Psi_{3/2^-}$ and (b) $\Psi_{1/2^+}$. 
The shaded region represents the measured cross section feeding the ground 
state of the residual core, $\sigma_{\rm exp}^{\rm gs}=3.3(1.5)$ ${\rm fm}^2$ 
\cite{exp2}. 
}
\label{netot} 
\end{figure}

In Fig. \ref{netot}, we present the sum of stripping and diffraction 
cross sections depending on $\beta_2$ and $E$. 
The sum of cross sections for the normalized deformed states [Eqs. (\ref{nestate1}) and (\ref{nestate2})] can be compared 
with the measured cross section feeding the ground state of the 
residual core, $\sigma_{\rm exp}^{\rm gs}=3.3(1.5)$ ${\rm fm}^2$ \cite{exp2}.  
$\sigma_{\rm str+diff}$ of $\Psi_{3/2^-}$ is comparable to 
$\sigma_{\rm exp}^{\rm gs}$ while that of $\Psi_{1/2^+}$ is larger than the 
measured one unless $\beta_2\lesssim0.3$. 

We expect that once the spectroscopic factors are specified, one can use 
the calculated cross sections and momentum distributions to deduce the spin-parity state of 
\mbox{$^{31}$Ne} and to determine the accurate values of $\beta_2$ and $E$. 
In Ref. \cite{exp2}, the experimental partial cross sections feeding the ground core state 
and excited core states have been determined. 
The neutron removal of the halo neutron from $^{31}\mbox{Ne}$ is expected 
to produce the core in its ground state. 
On the other hand, if the core is produced in an excited state, the removed neutron is likely to be one of the non-halo neutrons from the core, 
$^{30}\mbox{Ne}$. 
Indeed, the cross section for populating excited core states [$\sigma^{\rm inc}_{\rm exp}-\sigma^{\rm gs}_{\rm exp}=90(7)-33(15)$ mb] is similar to 
that of neutron removal from $^{30}$Ne [62(2) mb at 228 MeV/nucleon] \cite{exp3}. 

\begin{figure}
\includegraphics[width=0.45\textwidth]{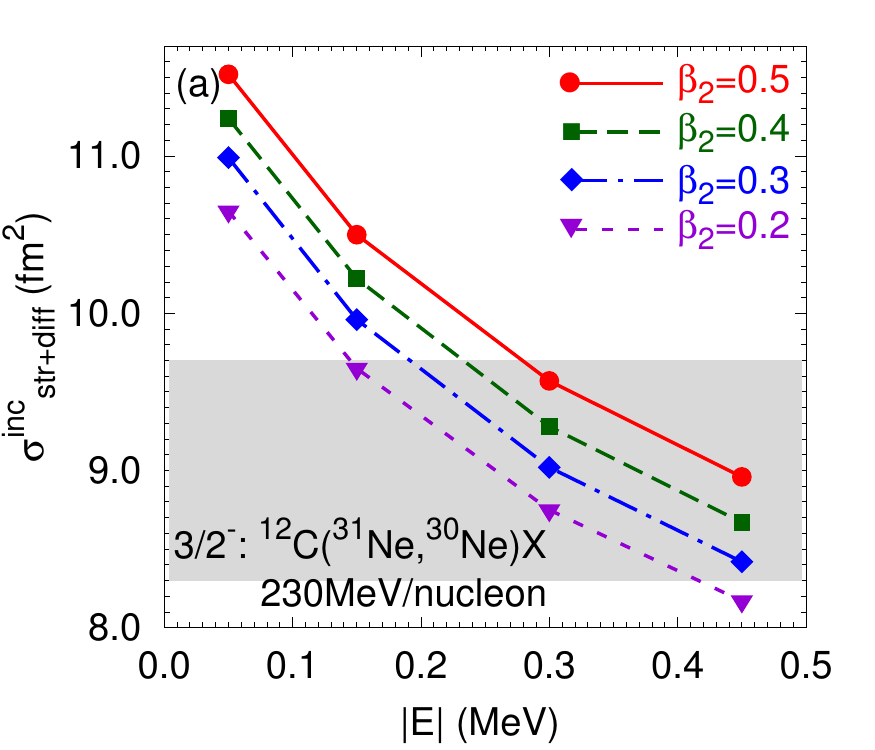}
\includegraphics[width=0.45\textwidth]{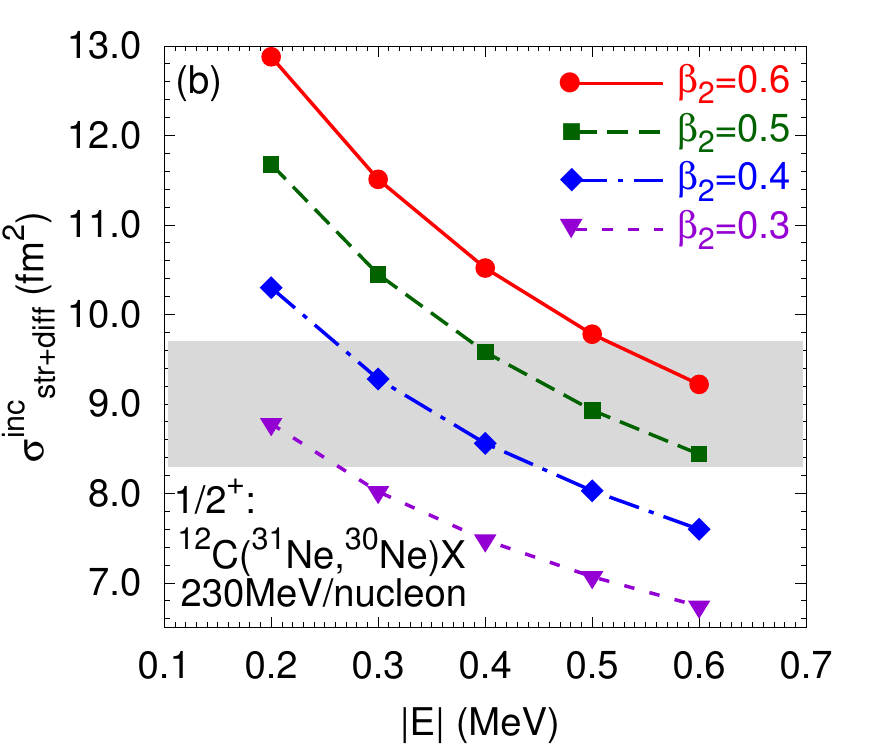}
\caption{ (Color online) 
The inclusive cross sections for the deformed 
states (a) $\Psi_{3/2^-}$ and (b) $\Psi_{1/2^+}$. 
The shaded region represents the experimental inclusive cross section, 
$\sigma_{\rm exp}^{\rm inc}=9.0(0.7)$ ${\rm fm}^2$ 
\cite{exp2}. 
}
\label{netot_exc} 
\end{figure}

Our calculations in the above are for knockout reactions of the halo neutron from $^{31}$Ne, populating the ground state of deformed core. 
For producing excited core states, we can consider neutron removal from the 
core, $^{30}$Ne. 
In order to compare with inclusive (populating both ground and excited core states) momentum distributions, we include spherical calculations for neutron removal from $^{30}$Ne and add them to the ground state calculations (see Fig. \ref{netot_exc}).  
For excited states of the residue, the neutron configurations of 
$2p_{3/2}$ and $1f_{7/2}$ are considered with the spectroscopic factors 
of 0.34 and 0.80 \cite{exp2}, respectively, in the case of the deformed state 
$\Psi_{3/2^-}$. For the state $\Psi_{1/2^+}$, $1d_{3/2}$ is considered with the spectroscopic 
factor 0.55 \cite{exp2}.
 
We mention that our approach has some limitations to analyze the cross sections populating excited core states.   
The cross sections of excited core states are independent of $\beta_2$ and their (especially $f$ and $d$ configurations) dependences on $E$ are relatively weaker 
than those of the ground core states. 
In addition, the cross sections are not exclusively determined by experimental measurements, and we can compare only the sum of cross sections for both ground 
and excited core states with data.
As a result, the comparison of the inclusive cross sections calculated in our method with experimental data is somewhat subtle. 

The comparisons of the ground state cross section (Fig. \ref{netot}) and the inclusive cross section (Fig. \ref{netot_exc}) with experimental data can be 
useful to investigate the possible ranges of $\beta_2$ and $E$. 
As discussed above, the dependences of the inclusive cross section on $\beta_2$ and $E$ are less clear though. 
For $0.3\lesssim\beta_2\lesssim0.5$, $\sigma^{\rm gs}_{\rm str+diff}$ and $\sigma^{\rm inc}_{\rm str+diff}$ of $\Psi_{3/2^-}$ agree with experimental data if 0.3 MeV$\lesssim|E|\lesssim0.45$ MeV. 
For $\beta_2\approx0.2$, the theoretical predictions with 0.15 MeV$\lesssim|E|\lesssim$ 0.3 MeV are comparable to the measured ones. 
On the other hand, $\sigma^{\rm gs}_{\rm str+diff}$ of $\Psi_{1/2}^+$ with strong deformation is larger, but $\sigma^{\rm inc}_{\rm str+diff}$ with weak deformation and strong binding is smaller than experimental data.   

\begin{figure}
\includegraphics[width=0.45\textwidth]{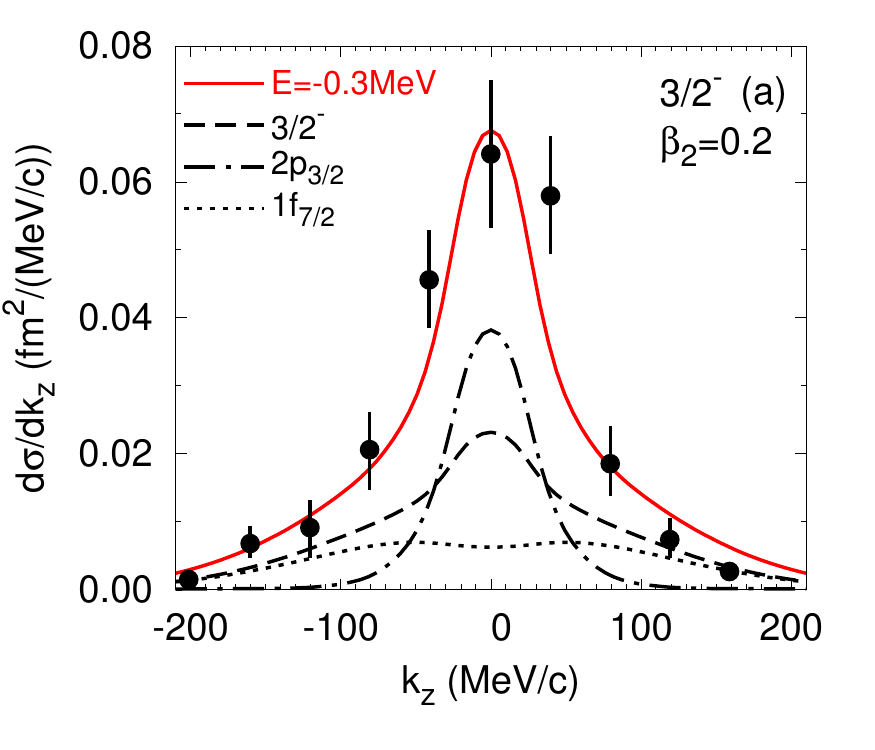}
\includegraphics[width=0.45\textwidth]{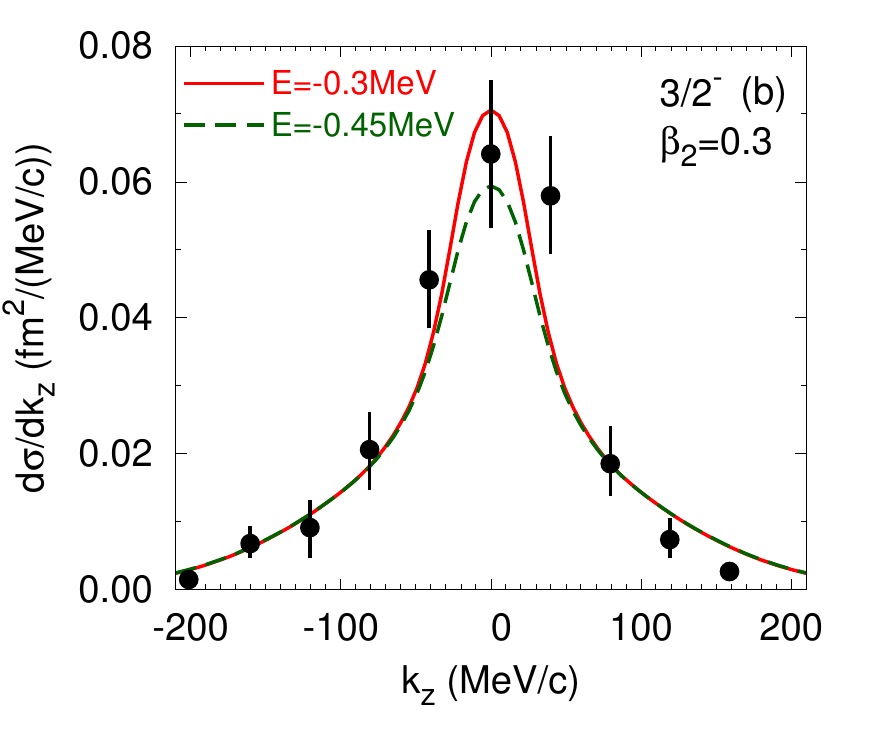}
\includegraphics[width=0.45\textwidth]{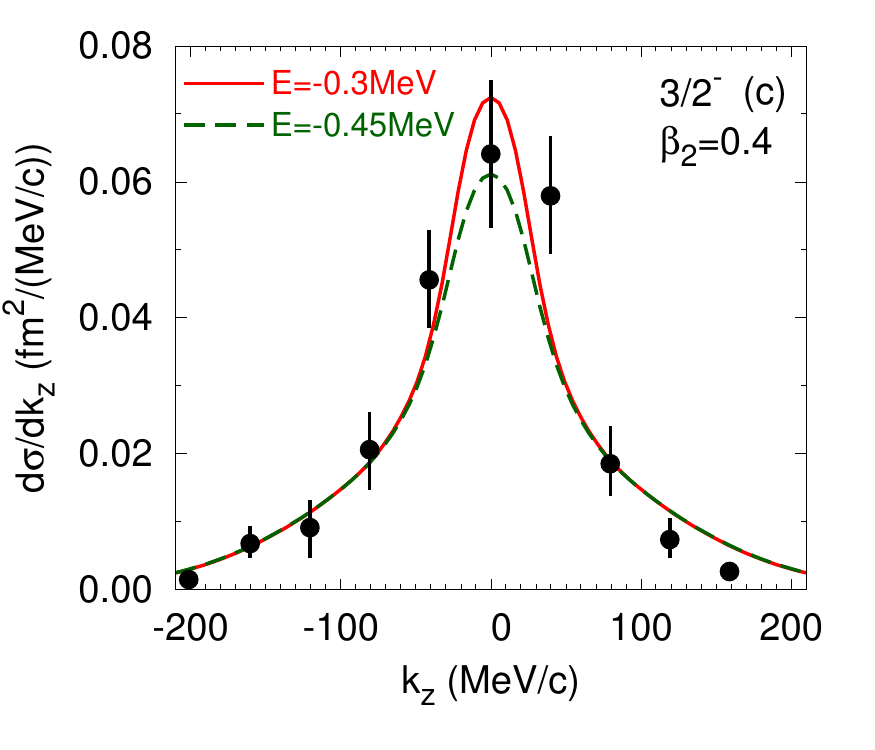}
\includegraphics[width=0.45\textwidth]{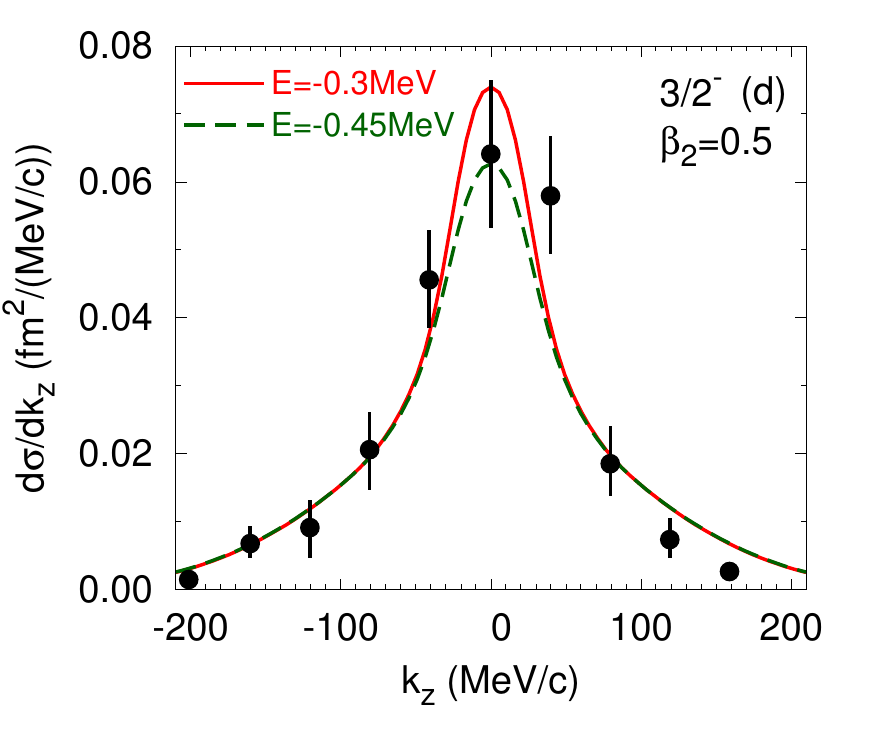}
\caption{ (Color online) 
The inclusive longitudinal momentum distribution for the deformed state $\Psi_{3/2^-}$. 
In (a), the valence neutron contributions from $\Psi_{3/2^-}$, $2p_{3/2}$, and 
$1f_{7/2}$ are shown as the dashed, dashed-dotted, and dotted line, 
respectively.  
The circles with error bars represent the inclusive one-neutron removal 
cross sections \cite{exp2}. 
}
\label{nemomdis} 
\end{figure}

\begin{figure}
\includegraphics[width=0.45\textwidth]{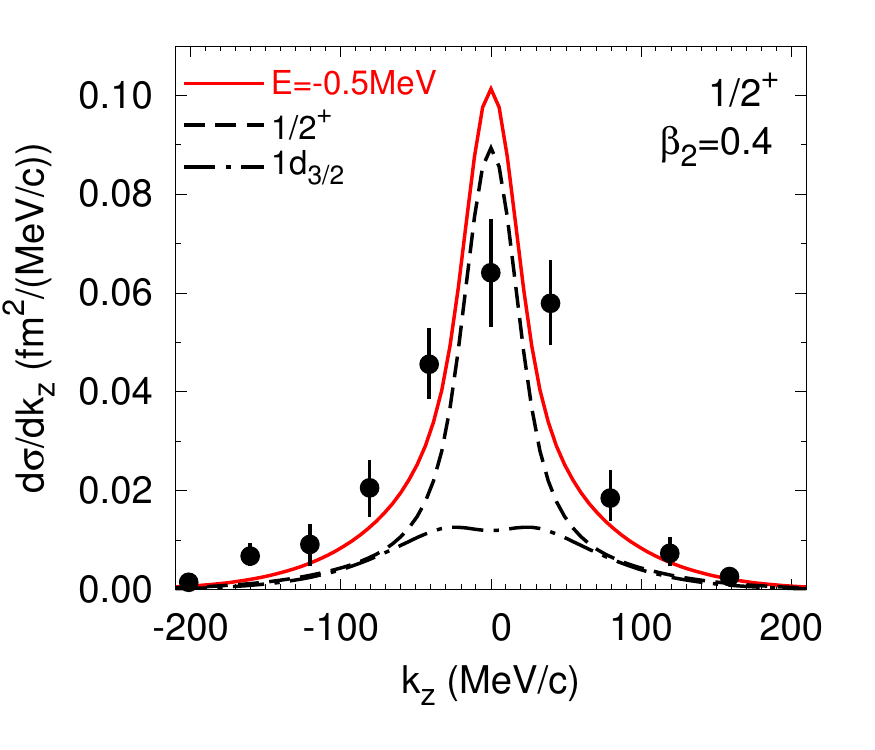}
\caption{ (Color online) 
Same as Fig. \ref{nemomdis} but 
for the deformed state $\Psi_{1/2^+}$. 
The valence neutron contributions from $\Psi_{1/2^+}$ and $1d_{3/2}$ are shown as 
the dashed and dashed-dotted lines, respectively. 
}
\label{negmomdis} 
\end{figure}

Figs. \ref{nemomdis} and \ref{negmomdis} show the comparison of 
the inclusive momentum distributions with experimental data. 
In Fig. \ref{nemomdis} (a), the partial cross sections feeding the ground ($3/2^-$) core 
state and excited ($2p_{3/2}$ and $1f_{7/2}$) core states are shown as styled lines. The solid line represents the sum of all the contributions and is compared with the data. 
For $\beta_2=0.2$, the inclusive momentum distribution with $E=-0.3$ MeV agrees with the experimental data.  
However, the momentum distribution with larger [smaller] $|E|$ is wider [narrower] than 
the measured one. 
For $0.3\lesssim\beta_2\lesssim0.5$, the momentum distributions are comparable if 0.3 MeV $\lesssim |E| \lesssim$ 0.45 MeV.  
The inclusive longitudinal momentum distributions are 
not very sensitive to deformation except near $k_z=0$. 
Especially, the results in Fig. \ref{nemomdis} (b), (c), and (d) exhibit 
almost same widths for each $E=-0.3, \, -0.45$ MeV.  
On the other hand, the longitudinal momentum distributions of the deformed state $\Psi_{1/2^+}$ do not agree with the 
experimental data. 
Fig. \ref{negmomdis} shows the momentum distribution with $\beta_2=0.4$, 
but for stronger deformation (which is suggested by Ref. \cite{hamamoto}) the distributions become narrower.

The calculated momentum distributions of the state $\Psi_{3/2^-}$ have the full 
width at half maximum (FWHM), approximately $82-93$ MeV/c, and they are narrower at low momentum and broader at high momentum than the measured one. 
The width of the measured momentum distribution is $77(18)$ MeV/c, extracted 
from a Lorentzian fit \cite{exp2}. 
In the contribution of the ground state of the residue, we have not included interference contributions which might 
account for the difference between our results and experimental data. 

For $\beta_2=0.5$ and $E=-0.15$ MeV, the probabilities for reactions through the  channels 
$p_{3/2}$, $f_{5/2}$, and $f_{7/2}$ in Fig. \ref{fgamp} (a) are close to those in Fig. 3 (b) of 
Ref. \cite{hamamoto}. 
The deformed state $\Psi_{3/2^-}$ seems to have a good correspondence with the Nilsson level 
[321 3/2]. 

\section{Summary}
\label{summary}

Using the Glauber model for knockout reactions, we have studied the
one-neutron removal reaction from the deformed projectile \mbox{$^{31}$Ne} incident on carbon targets at 230 MeV/nucleon. 
We have generated single particle wave functions with a deformed Woods-Saxon potential to calculate longitudinal 
momentum distributions using an orientation-dependent core-target S-matrix. 
The calculated longitudinal momentum distributions and cross sections have 
been analyzed with a quadrupole deformation parameter and an effective binding energy of 
the valence neutron. 

We observe that the cross section for the reaction increases with the wave function rms radius of the deformed 
states and has the inverse trend to the average $l$ value. This trend is meaningful based on the interpretation
of the role of the centrifugal barrier.
The width of the momentum distribution is also sensitive to the effective binding 
energy of the valence neutron, as it determines the extension of the single particle states. 

Our major conclusions are as follows. 
The sum of stripping and diffraction cross sections of the normalized state 
$\Psi_{3/2^-}$ is comparable with the measured cross section feeding the ground state of the residue. 
By including the neutron removal from the core with the neutron 
configurations $p$ and $f$, the inclusive momentum distribution and the total 
cross section for $\beta_2\approx0.2$ and $0.3\lesssim\beta_2\lesssim0.5$ agree with experimental data if $|E|\approx$ 0.3 MeV and 0.3 MeV $\lesssim|E|\lesssim$ 0.45 MeV, respectively. 
We mention that the inclusive longitudinal momentum distribution is not very sensitive to deformation 
at least for $\beta_2=0.3-0.5$. 
By including spherical calculations for excited core states, the dependences of 
the total cross sections and the longitudinal momentum distributions on $\beta_2$ and $E$ become weaker than those of the ground core states. 
In that respect, our approach has some subtleties to analyze the total inclusive cross sections. 
Our result, nevertheless, is consistent with the analysis of Refs. \cite{hamamoto} and 
\cite{exp2} in which a small neutron separation energy 
$S_n=0.15^{+0.16}_{-0.10}$ MeV is obtained (the measured one 
$S_n=0.29\pm1.64$ MeV \cite{NeSn} contains large uncertainties). 
On the other hand, the cross sections of $\Psi_{1/2^+}$ are larger and the 
widths 
of their corresponding momentum distributions are narrower than experimental 
data unless the core is weakly deformed (which disagrees with Ref. \cite{hamamoto}). 

Our results indicate that \mbox{$^{31}$Ne} has the spin parity $3/2^-$. 
With exclusive experimental measurements of cross sections and momentum distributions for both ground and excited core states, 
our method can be used further to study the spin-parity state of deformed 
nuclei, and the effects of deformation and binding energies, on nucleon removal 
reactions. 

\bigskip

\noindent{\bf ACKNOWLEDGMENTS}

This work is supported by the Rare Isotope Science Project of Institute for
Basic Science funded by Ministry of Science, ICT and Future Planning and
National Research Foundation of Korea (2013M7A1A1075764).  C.A.B. acknowledges 
support under U.S. DOE Grant DDE-FG02-08ER41533 and U.S. NSF grant PHY-1415656.
A.K. acknowledges funding support from Hungarian Scientific Research Fund OTKA K112962.

\appendix

\section{The Eikonal Phase for Quadrupole Deformed Core}

In the second term of Eq. (\ref{chidef}), 
\begin{eqnarray}
\frac{1}{2\pi}\int d^2\q \, \rho_t(q)f_{NN}(q) \, e^{-i(\b-\rrho)\cdot\q}
&=&\frac{1}{2\pi}
\int dq \, q \rho_t(q)f_{NN}(q) \int d\phi_q \, e^{-i|\b-\rrho|q\cos\phi_q} \, ,
\nonumber\\
&=&\int dq \, q\rho_t(q)f_{NN}(q)J_0(|\b-\rrho|q) \, .
\end{eqnarray}
Using the Graf's addition theorem \cite{AS64}, 
\begin{equation}
J_0(\sqrt{x^2+y^2-2xy\cos\phi})
=\sum_n J_n(x)J_n(y)e^{in\phi} \, ,
\end{equation}
we have 
\begin{eqnarray}
\frac{1}{2\pi}\int d^2\q \, \rho_t(q)f_{NN}(q) \, e^{-i(\b-\rrho)\cdot\q}
&=&\int dq \, q\rho_t(q)f_{NN}(q)\sum_n J_n(qb)J_n(q\rho)e^{in\phi} \, .
\end{eqnarray}
Then, for quadrupole deformed core, the second term of Eq. (\ref{chidef}) becomes 
\begin{eqnarray}
\label{chidefapp}
&&
\sum_m R_0\beta_2 D^2_{m0}(\hat{\OOmega})
\int d^3\r Y_{2m}(\hat{\r}) \frac{\partial \rho_c}{\partial r}\bigg\vert_{\beta_2=0}
\frac{1}{2\pi k_{NN}}\int d^2\q \rho_t(q)f_{NN}(q)e^{-i(\b-\rrho)\cdot\q}
\nonumber\\
&=&\sum_m \frac{1}{k_{NN}}R_0\beta_2 D^2_{m0}(\hat{\OOmega})
\int dr \, r^2 \int d\theta \, \sin\theta \int d\phi \, 
Y_{2m}(\hat{\r}) \frac{\partial\rho_c}{\partial r}\bigg\vert_{\beta_2=0}
\nonumber\\
&&\quad\times
\int dq \, q\rho_t(q)f_{NN}(q)\sum_n J_n(qb)J_n(q\rho)e^{in\phi} \, .
\nonumber\\
\end{eqnarray}
By integrating over $\phi$ and $\theta$, Eq. (\ref{chidefapp}) becomes 
\begin{eqnarray}
&&\sum_m \sqrt{\frac{(2-m)!}{(2+m)!}}\frac{\sqrt{5\pi}}{k_{NN}} R_0\beta_2 D^2_{m0}(\hat{\OOmega})
\int dr \, r^2 \int d\theta \, \sin\theta \,
P_{2}^m(\cos\theta) 
\frac{\partial\rho_c}{\partial r}\bigg\vert_{\beta_2=0}
\nonumber\\
&&\qquad\times
\int dq \, q\rho_t(q)f_{NN}(q) J_{-m}(qb)J_{-m}(qr\sin\theta)  
\nonumber\\
&=&
\frac{\sqrt{5\pi}}{k_{NN}} R_0\beta_2 D^2_{00}(\hat{\OOmega})
\int dr \,  \frac{\partial \rho_c}{ \partial r}\bigg\vert_{\beta_2=0}
\nonumber\\
&&\qquad\times
\int dq \, J_0(qb) \rho_t(q)f_{NN}(q) 
\frac{1}{q^2r}\Big[(3-q^2r^2)\sin(qr)-3qr\cos(qr)\Big]
 \nonumber\\
&&+\sqrt{\frac{15\pi}{2}}\frac{1}{k_{NN}} R_0\beta_2 
[D^2_{20}(\hat{\OOmega})+D^2_{-20}(\hat{\OOmega})]
\int dr \,  \frac{\partial \rho_c}{ \partial r}\bigg\vert_{\beta_2=0}
\nonumber\\
&&\qquad\times
\int dq \, J_2(qb) \rho_t(q)f_{NN}(q) 
\frac{1}{q^2r}\Big[(3-q^2r^2)\sin(qr)-3qr\cos(qr)\Big] \, .
\end{eqnarray}


\begin{thebibliography}{99}


\bibitem{exp1} T. Nakamura {\it et al.}, Phys. Rev. Lett. {\bf 103}, 262501 (2009).  

\bibitem{ne9} A. Poves and J. Retamosa, Nucl. Phys. A {\bf 571}, 221 (1994).  

\bibitem{ne10} P. Descouvemont, Nucl. Phys. A {\bf 655}, 440 (1999).  

\bibitem{ne16} K. Minomo {\it et al.}, Phys. Rev. C {\bf 84}, 034602 (2011).  

\bibitem{ne17} T. Sumi {\it et al.}, Phys. Rev. C {\bf 85}, 064613 (2012).  

\bibitem{ne18} K. Minomo {\it et al.}, Phys. Rev. Lett. {\bf 108}, 052503 (2012).  

\bibitem{ne22} W. Horiuchi, Y. Suzuki, P. Capel and D. Baye, Phys. Rev. C {\bf 81}, 024606 (2010).  

\bibitem{ne19} Y. Urata, K. Hagino and H. Sagawa, Phys. Rev. C {\bf 83}, 041303(R) (2011).  

\bibitem{ne20} Y. Urata, K. Hagino and H. Sagawa, Phys. Rev. C {\bf 86}, 044613 (2012).  

\bibitem{ne21} M. Takechi {\it et al.}, Phys. Lett. B {\bf 707}, 357 (2012).  

\bibitem{Ne} Shubhchintak and R. Chatterjee, Nucl. Phys. A {\bf 922}, 99 (2014). 

\bibitem{hamamoto} Ikuko Hamamoto, Phys. Rev. C {\bf 81}, 021304(R) (2010). 

\bibitem{bertsch} K. Hencken, G. Bertsch and H. Esbensen, Phys. Rev. C {\bf 54}, 3043 (1996).  

\bibitem{tostevin_hansen} P. G. Hansen and J. A. Tostevin, Annu. Rev. Nucl. Part. Sci. {\bf 53}, 219 (2003).  

\bibitem{momdis} C. A. Bertulani and A. Gade, Comp. Phys. Comm. {\bf 175}, 372 (2006).  

\bibitem{bertulani} C. A. Bertulani and P. G. Hansen, Phys. Rev. C {\bf 70}, 034609 (2004).  

\bibitem{zelevinsky} A. Sakharuk and V. Zelevinsky, Phys. Rev. C {\bf 61}, 014609 (1999).  

\bibitem{tostevin3} J. A. Christley and J. A. Tostevin, Phys. Rev. C {\bf 59}, 2309 (1999).  

\bibitem{tostevin2} P. Batham, I. J. Thompson and J. A. Tostevin, Phys. Rev. C {\bf 71}, 064608 (2005).  

\bibitem{tostevin} E. C. Simpson and J. A. Tostevin, Phys. Rev. C {\bf 86}, 054603 (2012).  

\bibitem{singh} G. Singh, Shubhchintak and R. Chatterjee, Phys. Rev. C 94, 024606 (2016).

\bibitem{esbensen} H. Esbensen, Phys. Rev. C {\bf 53}, 2007 (1996).  

\bibitem{book} C. A. Bertulani and P. Danielewicz, {\it Introduction to Nuclear Reactions}, (IOP Publishing, Bristol, UK, 2004).  

\bibitem{trr} M. S. Hussein, R. A. Rego and C. A. Bertulani, Phys. Rept. {\bf 201}, 279 (1991).  

\bibitem{fnn} L. Ray, Phys. Rev. C {\bf 20}, 1957 (1979).  

\bibitem{HM85} M. S. Hussein and K. W. McVoy, Nucl. Phys. A445 (1985) 124.

\bibitem{pseudo} A. T. Kruppa and Z. Papp, Comp. Phys. Comm. {\bf 36}, 59 (1985).  

\bibitem{defchi} P. J. Moffa, C. B. Dover and J. P. Vary, Phys. Rev. C {\bf 16}, 1857 (1977).  

\bibitem{My70} W.D. Myers, Nucl. Phys. A 145, 387 (1970).

\bibitem{BohrMottelson} A. Bohr and B. R. Mottelson, {\it Nuclear Structure} (Benjamin, Reading, MA, 1969), Vol. I. 

\bibitem{Gade2008} A. Gade {\it et al.}, Phys. Rev. C 77, 044306 (2008).

\bibitem{DeVries} H. De Vries, C. W. De Jager and C. De Vries, Atom. Data. Nucl. Data 36, 495 (1987).

\bibitem{exp2} T. Nakamura {\it et al.}, Phys. Rev. Lett. {\bf 112}, 142501 (2014).  

\bibitem{exp3} H. N. Liu {\it et al.}, Phys. Lett. B {\bf 767}, 58 (2017).  

\bibitem{NeSn} B. Jurado {\it et al.}, Phys. Lett. B {\bf 649}, 43 (2007).  

\bibitem{AS64} M. Abramowitz and I. A. Stegun, {\it Handbook of mathematical functions}, Dover Publications (1964). 


\end{thebibliography}
\end{document}